\documentclass[journal=jacsat,manuscript=article]{achemso}
\usepackage{xcolor}
\usepackage{multirow}
\usepackage{mathtools}
\usepackage{braket}
\usepackage{makecell}
\usepackage{graphicx}
\usepackage{dcolumn}
\usepackage{bm}
\usepackage{mathrsfs}   
\usepackage{tabularray}
\usepackage{makecell}
\usepackage{diagbox}
\usepackage{pdfpages}
\usepackage{comment}
\SectionNumbersOn
\newcommand{\nmax}{\ensuremath{q    }}
\newcommand{\nmaxCT}{\ensuremath{q_{\text{CT}}}}
\newcommand{\reddown}{\underline{\textcolor{red}{\downarrow}\phantom{\uparrow}}\;}
\newcommand{\redup}{\underline{\textcolor{red}{\uparrow}\phantom{\uparrow}}\;}
\newcommand{\bluedown}{\underline{\textcolor{blue}{\downarrow}\phantom{\uparrow}}\;}
\newcommand{\blueup}{\underline{\textcolor{blue}{\uparrow}\phantom{\uparrow}}\;}
\newcommand{\redocc}{\underline{\textcolor{red}{\uparrow\downarrow}}\;}

\definecolor{cadmiumgreen}{rgb}{0.0, 0.42, 0.24}

\newcommand{\burd}{\underline{\textcolor{blue}{\uparrow}\textcolor{red}{\downarrow}}\;}
\newcommand{\rubd}{\underline{\textcolor{red}{\uparrow}\textcolor{blue}{\downarrow}}\;}
\newcommand{\unocc}{\underline{\phantom{\uparrow}\phantom{\uparrow}}\;}
\DeclareMathOperator{\Tr}{Tr}
\newcolumntype{H}{>{\setbox0=\hbox\bgroup}c<{\egroup}@{}}
\title{Automatic State Interaction with Large Localized Active Spaces for Multimetallic Systems
}
\author{Valay Agarawal}
\author{Daniel King}
\author{Matthew R. Hermes}
\email{mrhermes@uchicago.edu}
\affiliation[UChicago]
{Department of Chemistry, University of Chicago.}
\author{Laura Gagliardi}
\email{lgagliardi@uchicago.edu}
\affiliation[UChicago]
{Department of Chemistry, University of Chicago.}

\begin{document}

\begin{abstract}
The localized active space self consistent field (LASSCF) method factorizes a complete active space (CAS) wave function into an antisymmetrized product of localized active space wave function fragments. Correlation between fragments is then reintroduced through LAS state interaction (LASSI), in which the Hamiltonian is diagonalized in a model space of LAS states. However, the optimal procedure for defining the LAS fragments and LASSI model space is unknown. We here present an automated framework to explore systematically convergent sets of model spaces, which we call LASSI[$r$,$\nmax$]. This method requires the user to select only $r$, the number of electron hops from one fragment to another and $\nmax$, the number of fragment basis functions per Hilbert space, which converges to CASCI in the limit of $r,\nmax\to\infty$. Numerical tests of this method on the tri-metal complexes [Fe(III)Al(III)Fe(II)($\mu_3$-O)]$^{6+}$ and [Fe(III)$_2$Fe(II)($\mu_3$-O)]$^{6+}$ show efficient convergence to the CASCI limit with 4-10 orders of magnitude fewer states.
\end{abstract}


\maketitle


\section{Introduction}
\label{sec:Introduction}
Transition metal systems pose a unique challenge for quantum chemistry due to the near-degeneracy of the $d$-shell electrons. These near-degeneracies mean that transition metal systems are often ill-described by any single electronic configuration, and thus often require multiconfigurational treatment to obtain accurate results. This is especially true in systems containing multiple transition metals with some degree of interaction. While multiconfigurational methods such as the complete active space self consistent field (CASSCF) method\cite{roos1980complete} are generally capable of describing a single metal quite well, they are often computationally prohibitive  when two or more metal centers are present.\cite{veryazov2011select,vogiatzis2017pushing}

However, a large number of configuration state functions (CSFs) (or Slater determinants (SDs) used in some codes) generated in the CAS method usually do not contribute in a substantial way to the final CAS wave function; these determinants are referred to as ``deadwood''\cite{ivanic2001identification}. 
Efforts to remove the deadwood from the wave function can be sorted broadly into two categories: ``automated'' strategies such as selected configuration interaction (SCI) \cite{Huron1973,Cimiraglia1987,Miralles1993,Neese2003,Tubman2016,Holmes2016} and density matrix renormalization group (DMRG) \cite{White1992} which incorporate the elimination of deadwood into the \textit{ansatz}; and ``manual'' strategies such as generalized active space SCF (GASSCF) \cite{gasscf1,gasscf2,gasscf3,gasscf4}, restricted active space SCF (RASSCF) \cite{rasscf,rasscf2}, the cluster mean field (cMF) \cite{Jimenez-Hoyos2015} and localized active space SCF (LASSCF) \cite{Hermes2019, hermes2020variational} methods, which utilize chemical knowledge to wisely eliminate deadwood prior to calculation. This latter category is clearly less black box, but comes with the opportunity to be exponentially more efficient prior to even beginning a calculation.

The LASSCF method (originally introduced as cMF \cite{Jimenez-Hoyos2015} but independently re-derived by us as a generalization of density matrix embedding theory \cite{Hermes2019,hermes2020variational}) is an approximation to CASSCF in which the wave function is factorized as a single antisymmetrized product of subspaces which are localized in real space. This approximation itself is not qualitatively accurate for multimetallic compounds involving interactions among the metal centers, even if they are weak, for the simple reason that the spin operator ($\hat{S}^2$) is nonlocal\cite{Pandharkar2019}. However, LASSCF states can serve as an efficient model space for the localized active space state interaction (LASSI) method,\cite{pandharkar2022localized} which recovers correlation between the subspaces. LASSI achieves this by projecting the molecular Hamiltonian into a model space of LAS wave functions which share the same set of orbitals in each subspace but possess different quantum numbers (e.g., number of electrons, local spin, and excitation number).


The LASSI method can be regarded as the application of the active space decomposition (ASD) method \cite{Parker2013,Parker2014,Parker2014a,Parker2014b,Kim2015} to the LASSCF reference wave function; or equivalently, a method intermediate between vector product state (VPS)-CASCI and VPS-CASSCF\cite{Nishio2022} (see also Ref.\ \citenum{Nishio2019}) in that orbitals are optimized, but only for the reference state. However, all investigations into these methods thus far have studied only organic chromophores for which the appropriate model space is known in advance to consist of singlet excitations, coupled triplet excited states, and single-electron charge transfer states. For systems such as multimetallic molecules -- key systems we would like to treat with these product-form methods -- the questions remain of what model spaces are appropriate, how best to explore different model spaces, and how quickly they can be expected to converge.




In previous applications of LASSI to transition metal systems, construction of the model space was carried out entirely ``manually'', i.e. the number of electrons and spin for each fragment in each state was decided by the user. This manual control facilitated a deeper understanding of the physics of the test systems; for instance the different contributions to the exchange coupling constant, $J$, for a simplified model\cite{pandharkar2022localized} of Kremer's Dimer\cite{sharma2020magnetic}. Nevertheless, for larger systems, this manual state selection process is time consuming, error prone, and non-scalable. 

In this paper, we propose a new method in which the LASSI model space selection scheme is ``automatic''. We call this method LASSI[$r$,$\nmax$]. In this approach, instead of selecting the individual configurations by hand, the user simply must select the number of ``electron hops'' from one fragment to another, $r$, and the number of locally excited states to construct for each fragment, $\nmax$. In the large-$r$, large-$\nmax$ limit, LASSI[$r$,$\nmax$] reproduces CASCI (including the exponential cost). We perform test calculations on a triiron complex [Fe(III)Fe(III)Fe(II)($\mu_3$-O)]$^{6+}$, previously explored with multireference methods in Ref.\ \citenum{vitillo2019quantum} and its aluminum substituted variant - [Fe(III)Al(III)Fe(II)($\mu_3$-O)]$^{6+}$ (henceforth ``Fe$_3$'' and ``AlFe$_2$''), and show that qualitative agreement with CASCI low-energy spectra is achieved by $r=1$ and $\nmax=15$ at the most, corresponding to multiple orders of magnitude fewer degrees of freedom than the CASCI limit. These calculations are described in Sec. \ref{sec:Results} below.

Through analysis of the LASSI eigenvectors of the Fe$_3$ system using a minimal (3$d$ only) active space, we furthermore identify more categories of LASSI model states which are deadwood, and eliminate them systematically through a further approximation which we call LASSI[$r$,$\nmaxCT$] (CT stands for charge transfer). This latter method, for the first time, permits us to carry out LASSI calculations for which the corresponding conventional CASCI is impossible: Fe$_3$ with a 3$d$ and 4$d$ active space. We contrast these results to those of a comparable DMRG calculation.

\section{Theory}
\label{sec:Theory}
Considering only the active-space part of the wave function (i.e., treating the doubly-occupied inactive electrons as part of the frozen core), the LAS wave function,
\begin{equation}
    \ket{\text{LAS}}=\bigwedge_K^{\text{fragments}} \ket{0_K}_{\mathcal{P}}, \label{eq:las_wfn}
\end{equation}
is a single antisymmetrized product of general many-electron fragment wave functions $\ket{0_K}_{\mathcal{P}}$, defined in a particular Hilbert space labeled $\mathcal{P}$, about which see below. The fragment wave functions are the eigenstates of the projected Hamiltonian,

\begin{eqnarray}\label{eqn:proj_hamiltonian}
    \hat{H}_{\mathcal{P}K} &\equiv& \hat{\mathcal{P}}_K^\dagger\hat{H}\hat{\mathcal{P}}_K,
\\ \label{eqn:projection_operator}
    \hat{\mathcal{P}}_K &=& \bigotimes_{L\neq K}^{\text{fragments}} \ket{0_L}_{\mathcal{P}}, \\
    0 &=& \left(\hat{H}_{\mathcal{P}K} - E_{n\mathcal{P}K}\right)\ket{n_K}_{\mathcal{P}}, \label{eq:model_state_eigenequation}
\end{eqnarray}

\noindent where $n \ge 0$ is an integer indexing the  excitation number of the particular fragment. Each fragment Hamiltonian, $\hat{H}_{\mathcal{P}K}$, describes the $K$th fragment in the mean field of all other fragments in their respective ground states. The LAS states in total are then indexed as 

\begin{equation}
\ket{\vec{n}}_{\mathcal{P}} \equiv \bigwedge_K^{\text{fragments}} \ket{n_K}_{\mathcal{P}}
\end{equation}

\noindent where $\vec{n}=\{n_1,n_2,n_3,\ldots\}$ is the ``excitation number vector'' (ENV) of nonnegative integers that index the various eigenstates of the $\hat{H}_{\mathcal{P}K}$ operators.

The Hilbert space indexed by $\mathcal{P}$ in which the roots of the projected Hamiltonians are sought (henceforth the `rootspace') is defined by the number of electrons in each fragment, the spin quantum numbers of each fragment, and the point group irreducible representation of each fragment wave function. In other words, $\mathcal{P}$ is a compound index representing the space table below, in which different rows represent different fragments and different columns represent different types of quantum number,
\begin{equation}
    \mathcal{P} = \begin{bmatrix}
    N_{\uparrow}^{(1)} & N_{\downarrow}^{(1)} & S^{(1)} & \Gamma^{(1)} \\
    N_{\uparrow}^{(2)} & N_{\downarrow}^{(2)} & S^{(2)} & \Gamma^{(2)} \\
    N_{\uparrow}^{(3)} & N_{\downarrow}^{(3)} & S^{(3)} & \Gamma^{(3)} \\
    \vdots & \vdots & \vdots & \vdots
    \end{bmatrix},
    \label{eq:rootspace}
\end{equation}
with $N_{\uparrow}$ for spin-up electrons, $N_{\downarrow}$ for spin-down electrons, $S$ for spin magnitude, and $\Gamma$ for point-group irrep. These fictitious, fragment-indexed quantum numbers are not conserved by the exact, correlated many-electron wave function of the whole system, but they are (in practice) conserved by the approximate $\ket{\text{LAS}}$ wave function. Much like a single determinant, a single LAS state is generally not an eigenstate of $\hat{S}^2$ unless at most one fragment has a nonzero $S^{(K)}$.

In addition to breaking spin symmetry, a single LAS state omits all entanglement between fragments; fragments interact exclusively \emph{via} a mean-field potential. This approximation is obviously quantitatively inaccurate, and can also be qualitatively inaccurate in certain cases \cite{Hermes2019}. However, interaction and entanglement between fragments, as well as global $\hat{S}^2$ symmetry, can be restored by constructing a linear combination of multiple states of the LAS type, which we refer to as LASSI \cite{pandharkar2022localized}. The LASSI wave functions are linear combinations of the LAS states $\ket{\vec{n}}_{\mathcal{P}}$:

\begin{equation}
    \ket{\text{LASSI}_i} \equiv \sum_{\mathcal{P}}^{\text{rootspaces}}\sum_{\vec{n}}^{\text{ENVs}}\ket{\vec{n}}_{\mathcal{P}}C^{(i)}_{\mathcal{P}\vec{n}}\label{eq:lassi_wfn}
\end{equation}

\noindent where $C^{(i)}_{\mathcal{P}\vec{n}}$ is an eigenvector coefficient of the Hamiltonian projected into the model space of LAS states:

\begin{equation}
    \braket{\text{LASSI}_i|\hat{H}|\vec{n}}_{\mathcal{P}}=E_{\text{LASSI}}^{(i)}C^{(i)}_{\mathcal{P}\vec{n}},
    \label{eq:lassi_schrodinger}
\end{equation}


\subsection{Analyzing LASSI results}

In Sec. \ref{sec:Results} below, we draw a distinction between significant model states and ``deadwood'' model states by analyzing the LASSI eigenvector, $C^{(i)}_{\mathcal{P}\vec{n}}$, which can be decomposed as
\begin{equation}
    C^{(i)}_{\mathcal{P}\vec{n}} = \sqrt{w_{\mathcal{P}}^{(i)}}u^{(i),\mathcal{P}}_{\vec{n}},
\end{equation}
with the normalized factors
\begin{eqnarray}
\sum_{\mathcal{P}}^{\text{rootspaces}} w_{\mathcal{P}}^{(i)} &=& 1\ \forall\ i \\
\sum_{\vec{n}}^{\text{ENVs}}(u^{(i),\mathcal{P}}_{\vec{n}})* u^{(i),\mathcal{P}}_{\vec{n}} &=& 1\ \forall\ i,\mathcal{P}.
\end{eqnarray}
Weights ($w_{\mathcal{P}}^{(i)}$) close to zero indicate that an entire rootspace is deadwood in the $i$th LASSI eigenstate. For rootspaces with nonzero weights, the ENV coefficients ($u^{(i),\mathcal{P}}_{\vec{n}}$) can be used to compute the single-fragment density matrix in the excitation number basis,
\begin{eqnarray}
    d_{m,m'}^{(i),K\mathcal{P}} &\equiv& \sum_{\vec{n}\in\{n_K=m\}}^{\text{ENVs}} \sum_{\vec{n}'\in\{n_K=m'\}}^{\text{ENVs}}\prod_{L\neq K}^{\text{fragments}}\delta_{n_Ln'_L}\nonumber \\ && \times(u^{(i),\mathcal{P}}_{\vec{n}})^* u^{(i),\mathcal{P}}_{\vec{n}'}.
\end{eqnarray}

In the application considered below, we examine spin-state energy gaps as obtained from energy differences of different LASSI eigenstates, and for this purpose, the weights and density matrices are averaged over the lowest-energy $i_{\textrm{max}}+1$ LASSI eigenstates,
\begin{eqnarray}
    w_\mathcal{P} &\equiv& \frac{1}{i_{\textrm{max}}+1}\sum_{i=0}^{i_{\textrm{max}}} w_\mathcal{P}^{(i)}, \label{eq:sa_weights}\\
    d_{m,m'}^{K\mathcal{P}} &\equiv& \frac{1}{i_{\textrm{max}}+1}\sum_{i=0}^{i_{\textrm{max}}} d_{m,m'}^{(i),K\mathcal{P}},
\end{eqnarray}
where in the spin-ladder applications below, $i_{\textrm{max}}+1$ is the number of states in the spin ladder.

From the (state-averaged) single-fragment density matrix, we extract the average excitation number of the $K$th fragment in the $\mathcal{P}$th rootspace,
\begin{equation}
\bar{n}_{K\mathcal{P}} = \sum_{m=0} m\times d^{K\mathcal{P}}_{m,m}, \label{eq:navg}
\end{equation}
as well as the von Neumann entropy,
\begin{eqnarray}
    s_{K\mathcal{P}} = -\sum_m\left( \rho_m^{K\mathcal{P}} \log \rho_m^{K\mathcal{P}}\right), \label{eq:vonneumanentropy}
\end{eqnarray}
where $\{\rho_m^{K\mathcal{P}}\}$ are the eigenvalues of the matrix $d^{K\mathcal{P}}$. If $\bar{n}_{K\mathcal{P}}$ and $s_{K\mathcal{P}}$ are close to zero, then excited states of $K$th fragment in the $\mathcal{P}$th rootspace are deadwood.

\subsection{LASSI[$r,\nmax$] \label{sec:model_space_selection}}

If the LASSI model space consists of all possible rootspaces described by all possible space tables [Eq.\ (\ref{eq:rootspace})] along with all possible excitation number vectors, then LASSI reproduces CASCI, with the same exponential-scaling computational cost. However, the conjecture underlying LAS methods in general (that electrons interact more strongly within fragments than between fragments) suggests that a small approximate model space could reproduce CASCI qualitatively with orders of magnitude fewer model states than the number of determinants or CSFs in the CAS. In order for this to be practical, such an approximate model space must be generable semi-automatically. As mentioned above, when LASSI was introduced in Ref.\ \citenum{pandharkar2022localized}, model states were constructed by the user entering all relevant space tables by hand; additionally, only the ground-state ENV ($\vec{0}=\{0_1,0_2,0_3,\ldots\}$) was considered in any rootspace. Here we explore model space construction more systematically.

First, we stipulate that all model spaces should be chosen to restore spin symmetry: LASSI wave functions should be $\hat{S}^2$ eigenvalues. This is accomplished by permuting
\begin{equation}
    M^{(K)} \equiv \frac{1}{2}\left(N^{(K)}_{\uparrow} - N^{(K)}_{\downarrow}\right)
\end{equation}
for each fragment in all possible ways consistent with the fixed whole-molecule total number of spin up and spin down electrons, as well as the constraint $M^{(K)}\le |S^{(K)}|$, and including all such permutations in any model space. Omitting point-group symmetry for the current work, the space table [Eq.\ (\ref{eq:rootspace})] can now be simplified to
\begin{equation}
    \mathcal{P} = \begin{bmatrix}
    N^{(1)} & S^{(1)} \\
    N^{(2)} & S^{(2)} \\
    N^{(3)} & S^{(3)} \\
    \vdots & \vdots
    \end{bmatrix},
    \label{eq:rootspace_simple}
\end{equation}
with the understanding that implicitly, all possible orientations of nonzero fragment spins are included in the model space.

Second, we construct simplified space tables [Eq.\ (\ref{eq:rootspace_simple})] by systematically moving one electron from any one fragment to any other fragment up to $r$ times in all possible ways, where $r$ is a non-negative integer chosen by the user. For each increment or decrement of $N^{(K)}$, the corresponding local spin $S^{(K)}$ can (in general) either increase or decrease by $\frac{1}{2}$. A schematic for this is presented in Table \ref{tab:rootspaces}.

\begin{table}[]
    \centering
    \begin{tabular}{|ccccc|c|}
    \hline
        \multicolumn{4}{|c}{Rootspace depiction}&\multicolumn{2}{c|}{\raisebox{-.5\height}{\includegraphics[width=20mm]{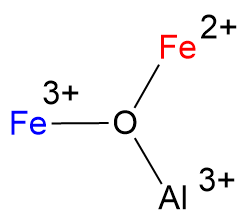}}}\\ \hline
       \multicolumn{6}{|c|}{\textbf{$r=0$}}\\
       \multicolumn{6}{|c|}{}\\
       (a) && $\redocc\redup\redup\redup\redup$ &&$\bluedown\bluedown\bluedown\bluedown\bluedown$ & \multirow{5}{*}{$\mathcal{P}=\begin{bmatrix} N^{(1)} & S^{(1)} \\ N^{(2)} & S^{(2)} \end{bmatrix}$} \\
       (b) && $\redocc\redup\redup\redup\reddown$ && $\bluedown\bluedown\bluedown\bluedown\blueup$ &\\
       (c) && $\redocc\redup\redup\reddown\reddown$ && $\bluedown\bluedown\bluedown\blueup\blueup$ &\\
       (d) && $\redocc\redup\reddown\reddown\reddown$ && $\bluedown\bluedown\blueup\blueup\blueup$ &\\
       (e) && $\redocc\reddown\reddown\reddown\reddown$ && $\bluedown\blueup\blueup\blueup\blueup$ &\\
       &&&&&\\\hline 
       \multicolumn{6}{|c|}{$r=1$}\\
       \multicolumn{6}{|c|}{}\\
       (f) && $\redup\redup\redup\redup\reddown$ && $\bluedown\bluedown\bluedown\bluedown\burd$& \multirow{2}{*}{$\begin{bmatrix} N^{(1)}-1 & S^{(1)}\pm 0.5 \\ N^{(2)}+1 & S^{(2)}-0.5 \end{bmatrix}$} \\
       (g) && $\redocc\redup\redup\redup\unocc$ && $\bluedown\bluedown\bluedown\bluedown\burd$& \\ 
       &&&&&\\
       (h)&&$\redocc\rubd\redup\redup\redup$&&$\bluedown\bluedown\bluedown\bluedown\unocc$&$\begin{bmatrix}N^{(1)}+1&S^{(1)}-0.5\\N^{(2)}-1&S^{(2)}-0.5 \end{bmatrix}$\\ \hline
       \multicolumn{6}{|c|}{$r=2$}\\
       \multicolumn{6}{|c|}{}\\
        (i) && $\rubd \redup\redup\redup\reddown$ && $\unocc\bluedown\bluedown\bluedown\burd$&$\begin{bmatrix}N^{(1)}&S^{(1)}\\ N^{(2)}&S^{(2)}-1\end{bmatrix}$ \\
        &&&&&\\
        (j) && $\redocc\rubd\redup\redup\unocc$ && $\unocc\bluedown\bluedown\bluedown\burd$&$\begin{bmatrix}N^{(1)}&S^{(1)}-1\\ N^{(2)}&S^{(2)}-1\end{bmatrix}$\\
        &&&&&\\
        (k) && $\redocc\rubd\rubd\redup\redup$ && $\bluedown\bluedown\bluedown\unocc\unocc$ &$\begin{bmatrix}N^{(1)}+2&S^{(1)}-1\\ N^{(2)}-2&S^{(2)}-1\end{bmatrix}$ \\\hline
    \end{tabular}
    \caption{Examples of the rootspaces included in a LASSI calculation for a LASSCF reference with the red Fe$^{2+}$ (6e,5o) atom and blue Fe$^{3+}$ (5e,5o) atoms, depicted in terms of representative single determinants, the corresponding simplified space table [Eq.\ \ref{eq:rootspace_simple}], and the $r$ value at which they appear for AlFe$_2$ systems. (a)--(e) are the various cases of the local spin polarizations, $2M^{(K)}\equiv N^{(K)}_\uparrow-N^{(K)}_\downarrow$. (f)--(h) are examples of $r=1$ rootspaces generated from (b) and are not exhaustive of all possible spin polarizations. Rootspace (f) adds 0.5 to $S^{(1)}$, and (g) decreases $S^{(1)}$ by 0.5 since $S$ corresponds to the number of singly occupied molecular orbitals divided by 2. (i)--(k) are examples of $r=2$ rootspaces. Specifically, (i), (j) and (k) rootspaces are generated by moving electron from the blue atom to the red atom from (f), (g) and (h) rootspaces respectively. $r=2$ rootspaces are not exhaustive.}
    \label{tab:rootspaces}
\end{table}

Third, we stipulate a maximum number of fragment roots $\nmax$, obtain up to the $\nmax$th eigenstate of each fragment Hamiltonian in each rootspace $\hat{H}_{\mathcal{P}K}$ (or all eigenstates if there are fewer than $\nmax$), and build all possible ENVs from this product space of these sets.

We use the formulation ``LASSI[$r,\nmax$]'' to refer to LASSI calculations whose model spaces are constructed in this way. Note that the CASCI limit is achieved as both parameters approach infinity:
\begin{equation}
    \lim_{r\to \infty} \lim_{\nmax \to \infty} \text{LASSI} [r,\nmax] \equiv \text{CASCI}.
\end{equation}

\subsection{Efficient Selection with LASSI[$r,\nmaxCT$] }
\label{sec:nmaxCT}

We will show evidence in Sec.\ \ref{sec:Results} from the results of LASSI[$r$,$\nmax$] calculations that it is profitable to use different number of eigenstates for different fragments in different rootspaces, because whole categories of model states can be shown to be ``deadwood'' by examining the LASSI eigenvectors. In particular, we will show that to obtain qualitative agreement with CASCI, it is often not necessary to include ``local excitations'' (i.e., $n_K>0$), \emph{except} where a particular fragment has a different number of electrons than in the reference LAS wave function. 

For instance, for a molecule with three fragments, if the reference ($r=0$) space table is written as Eq.\ (\ref{eq:rootspace_simple}), so that a single $r=1$ space ($\mathcal{Q}$) table might be
\begin{equation}
    \mathcal{Q} = \begin{bmatrix}
    N^{(1)}+1 & S^{(1)}+\frac{1}{2} \\
    N^{(2)}-1 & S^{(2)}+\frac{1}{2} \\
    N^{(3)} & S^{(3)}
    \end{bmatrix},    \label{eq:rootspace_r1}
\end{equation}
then only in the first and second fragments of this rootspace do local excited states, $n_K>0$, contribute to low-energy LASSI eigenstates.

We will therefore define ``LASSI[$r,\nmaxCT$]'' as an efficient approximate alternative to LASSI[$r,\nmax$], where ``CT'' stands for ``charge transfer'' and indicates that the model space includes up to the $\nmax$th local eigenstate \emph{only} if the number of electrons in the given fragment in the given rootspace differs from that in the reference rootspace; otherwise, only the ground eigenstate of $\hat{H}_{\mathcal{P}K}$ is included. Such calculations are indicated in the text by the Roman subscript CT on the integer $\nmax$ value; for instance ``LASSI[$1,5_{\text{CT}}$].''

\section{Computational Methods}
\begin{figure*}
\centering
\includegraphics[width=100mm]{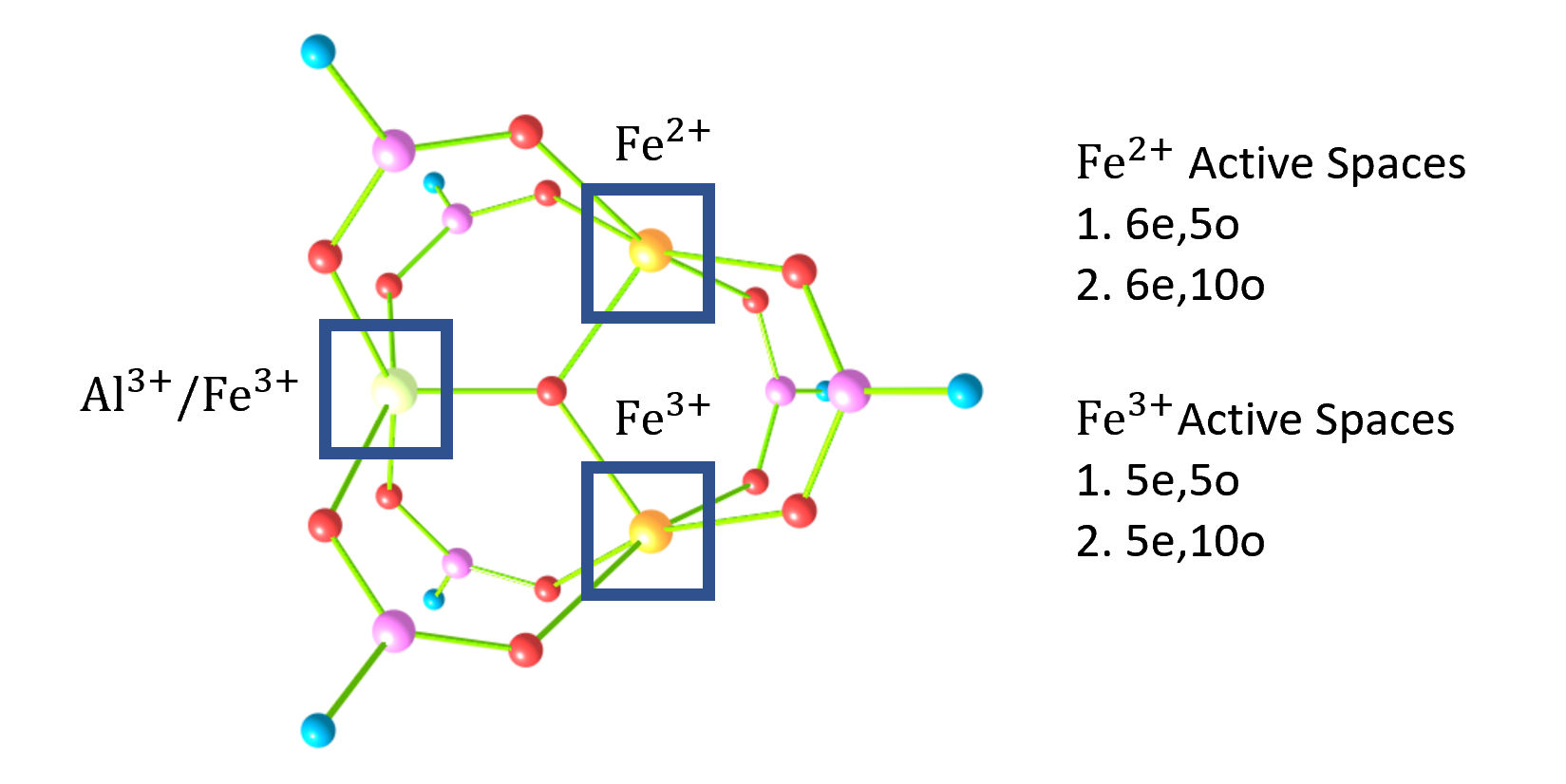}
\caption{AlFe$_2$ and Fe$_3$ system under study. Yellow atoms are Fe, red atoms are O, pink atoms are C, blue atoms are H. The green atom is Al for AlFe$_2$ system and Fe for Fe$_3$ system. All active orbitals are $3d$ and $4d$ iron orbitals. All Fe$^{2+}$ fragments were assigned a spin magnitude of $S=2$ (quintet) and all Fe$^{3+}$ fragments were assigned a spin magnitude of $S=\frac{5}{2}$ (sextet) in the reference LASSCF wave function. When optimizing orbitals using ROHF or LASSCF, the AlFe$_2$ system was fixed at $M\equiv\frac{1}{2}(N_\uparrow-N_\downarrow)=\frac{9}{2}$ and the Fe$_3$ system was fixed at $M=7$. When carrying out LASSI calculations, AlFe$_2$ was initialized at $M=\frac{1}{2}$ and Fe$_3$ was initialized at $M=0$. }
\label{fig:systems}
\end{figure*}

In this work, we studied two compounds reported in Ref. \citenum{vitillo2019quantum}: [Fe(III)Al(III)Fe(II)($\mu_3$-O)]$^{6+}$  and [Fe(III)$_2$Fe(II)($\mu_3$-O)]$^{6+}$. They are interesting chemical species because they are the building blocks in several metal-organic frameworks. The first system is a simplified model due to replacement of one Fe$^{3+}$ atom with a closed shell Al$^{3+}$, reducing the number of active orbitals that need to be considered. The geometries of the systems were taken from Ref. \citenum{vitillo2019quantum} without further optimization. The structure and active spaces of the systems under study are depicted in Fig.\ \ref{fig:systems}. To generate an initial guess for all MR calculations, we perform a high-spin ($S=9/2$ for AlFe$_2$ or $S=7$ for Fe$_3$) restricted open shell Hartree-Fock (ROHF) calculation, followed by atomic valence active space (AVAS) localization\cite{AVAS} for 3$d$ and 4$d$ orbitals of all Fe atoms. Note that AVAS generates larger-than-intended active spaces, and required active spaces are hand-picked to generate the initial guess for all calculations. At the LASSCF level, the orbitals are optimized for the high-spin state ($S=9/2$ for AlFe$_2$ system and $S=7$ for Fe$_3$ system), and these optimized LASSCF orbitals are used in all subsequent LASSI and CASCI calculations. Below, we refer to the CASCI results as LAS-CASCI. 

We used the cc-pVDZ \cite{dunning1989gaussian} basis set for all C and H atoms, and cc-pVTZ \cite{dunning1989gaussian} for all Al, Fe and O atoms. Since we use both 3$d$ and 4$d$ orbitals for Fe atoms, we need at-least triple $\zeta$ basis set for all Fe atoms. Additionally, we used cc-pVTZ for Al and O atoms since previous studies show contribution from O atoms even when orbitals on O are not taken into active space \cite{vitillo2019quantum}. The H and C atoms are modeled with relatively smaller basis set (cc-pVDZ) since they do not have contributions to the active space and it helps with reducing cost. LASSCF and LASSI are implemented in \textit{mrh} \cite{mrh_software}. CASSCF and CASCI calculations were performed using PySCF \cite{sun2020recent}.

\section{Results and discussion}
\label{sec:Results}
\subsection{Aluminium Di Iron system}
This system consists of one $d^6$ Fe$^{2+}$center, one $d^5$ Fe$^{3+}$ center and a $d^0$ Al$^{3+}$ center. Since Al$^{3+}$ is closed shell, we do not consider any active space on it. We explore two different active spaces for this system as shown in Fig.\ \ref{fig:systems}, namely (11e,10o) with all 3$d$ orbitals of both Fe atoms and (11e,20o) with all 3$d$ and 4$d$ orbitals of both Fe atoms. In these different parameters, we investigate the dependence of the $J$ coupling according to the Yamaguchi formula\cite{yamaguchi1986molecular}
\begin{equation}
    J = -\frac{E_{HS} - E_{LS}}{\braket{S^2}_{HS} - \braket{S^2}_{LS}} \label{eq:yamaguchi}
\end{equation} for the sake of brevity. The full spin ladder is reported in the SI table II and III.

\begin{figure}
    \centering
    \includegraphics[width=\linewidth]{ 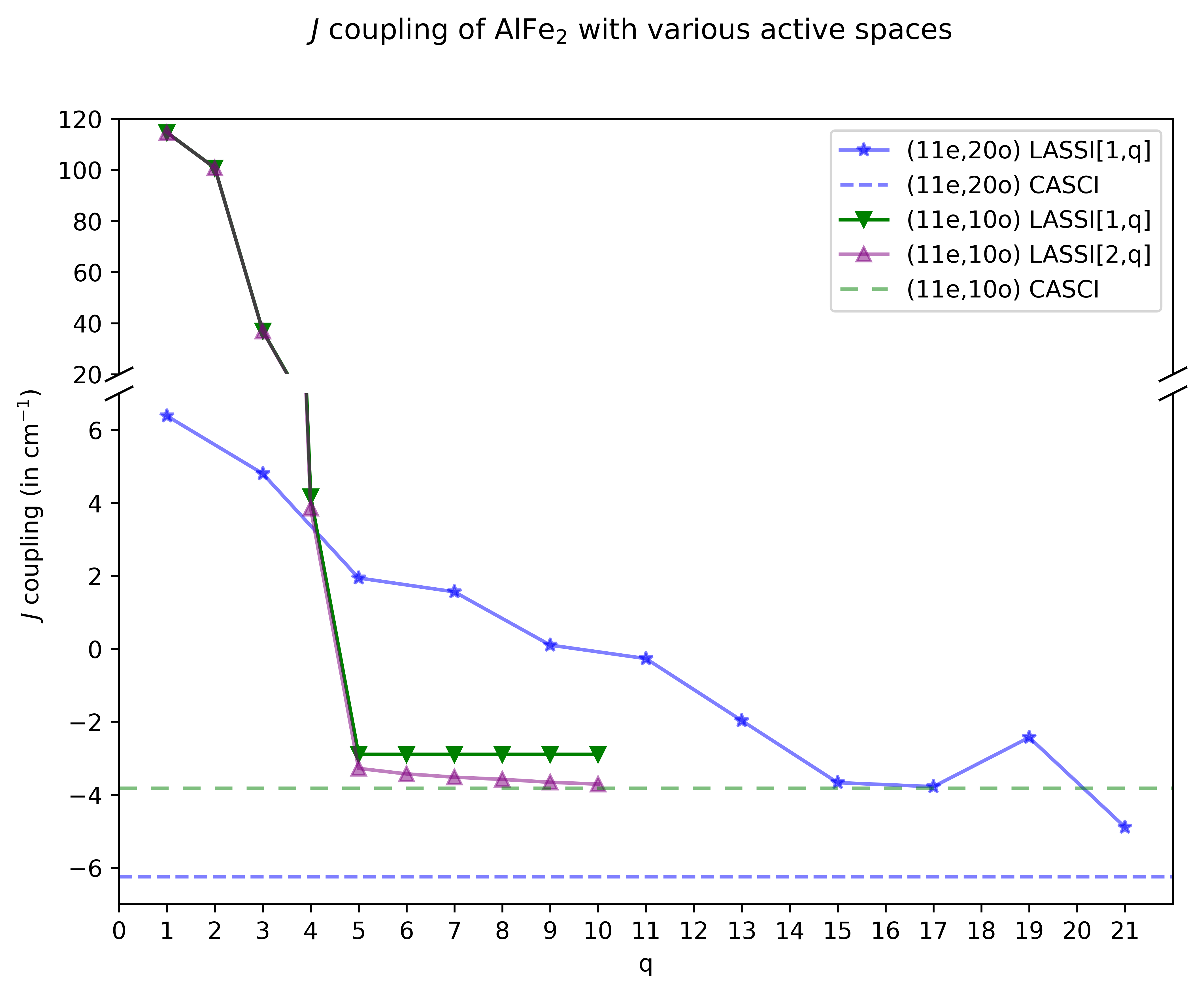}
    \caption{LASSI[1,$\nmax$] and CASCI predicted $J$ coupling for an AlFe$_2$ system with active spaces of (11e,10), and (11e,20o). For (11e,10o) active space, $J_{\text{LASSI}[0,\nmax]}=+116.69cm^{-1}\; \forall\; \nmax=\{1,10\}$ and $J_{\text{LASSI}[2,\nmax]} \approx J_{\text{LASSI}[3,\nmax]} \;\forall\; \nmax=\{1,10\}$.The LAS-CASCI result for(11e,10o) is -3.82 cm$^{-1}$ and for (11e,20o) is -6.245 cm$^{-1}$. }
    \label{fig:AlFe2_20_11}
\end{figure}

In Fig. \ref{fig:AlFe2_20_11}, we compare the performance of LASSI with its theoretical limit LAS-CASCI, under two variations. We first vary $r$ and $\nmax$ for the  (11e,10o) active space. LASSI[0,1] includes no charge transfer rootspaces and only ground state eigenfunctions of the projected Hamiltonian in each fragment (equivalent to (a)--(e) rootspaces in table \ref{tab:rootspaces}). This generates only enough model states to ensure that the LASSI eigenstates are eigenfunctions of the $\hat{S}^2$ operator and wrongly predicts ferromagnetic interaction ($J>0$) as compared to LAS-CASCI due to the complete omission of kinetic exchange. As $r$ and $\nmax$ increase, the agreement between LASSI and LAS-CASCI improves. The correct negative sign of $J$ is observed for $r>0$; $\nmax > 4$. At LASSI[1,5] (Fig. \ref{fig:AlFe2_20_11}: green curve with upside down triangles), we observe agreement within CASCI to within 1 cm$^{-1}$, from a wave function built from 250 model states. At LASSI[2,10] (1780 states) (purple curve with triangles), the difference is less than 0.1 cm$^{-1}$, at the price of requiring about 7 times as many model states compared to LASSI[1,5]. No performance improvement was observed for LASSI[3,$\nmax$] series. LASSI[2,10] wave function still has another 30 times fewer model states than the 52920 single determinants in the CASCI limit. All corresponding absolute energies are in SI table II and III. For comparison, CASSCF predicts a $J$ coupling of -7.3 cm$^{-1}$, as opposed to the LAS-CASCI limit of -3.8 cm$^{-1}$, which suggests that any inaccuracy in the LASSI[1,5] result is more likely attributable to the LASSCF orbitals than to the LASSI approximation. 

The second variation explored is the active space dependence of $J$. For all three active space choices, agreement of $J$ to within 3 cm$^{-1}$ was achieved with vastly fewer model states than the number of single determinants in the corresponding CASCI: 7425 model states for the LASSI[1,15](11e,20o) calculation, as compared to $\approx 6\times 10^8$ determinants in the CASCI(11e,20o) wave function. Clearly, the convergence with respect to $\nmax$ is less rapid in the larger active space. However, we note that the von Neumann entropies [Eq.\ (\ref{eq:vonneumanentropy})] for the LASSI[1,15](11e,20o) ($\text{max}(s_{K\mathcal{P}}) =0.95$) results are similar to the LASSI[1,5](11e,10o) ($\text{max}(s_{K\mathcal{P}}) =0.95$) results. (See supporting information for a complete list.) This means that the entanglement between the two fragments does not increase in the larger active space, suggesting that the higher required $\nmax$ in the larger active space is due rather to the less-accurate modeling of each fragment individually by the LASSI model states constructed from the eigenstates of $\hat{H}_{\mathcal{P}K}$.

\subsection{Tri-Iron system \label{sec:fe3_results}}
This system consists of one $d^6$ Fe$^{2+}$ center and two $d^5$ Fe$^{3+}$ centers. We explore the (16e,15o) active space of 3$d$ orbitals as well as the (16e,30o) active space including 3$d$ and 4$d$ orbitals. This system is more challenging for MR methods due to the size of active space, and as a reminder, DFT predicts that all Fe centers are identical\cite{vitillo2019quantum}. We calculate $S=0, 1, ..., 7$ spin states for each active space for all methods.

\begin{table}[]
\centering
\begin{tabular}{c|c|c|c|c|c}
\multicolumn{6}{c}{Spin Ladder (in kJ/mol)}\\\hline
Spin & CASPT2$^\dagger$&CASSCF$^\dagger$ & LAS-CASCI & LASSI[1,5]& LASSI[1,5$_\text{CT}$] \\\hline
$S=7$ & 22.8 & 6.6    & 5.0       & 4.9& 4.9    \\
$S=6$ & 21.0 & 4.8    & 3.5       & 3.4& 3.4    \\
$S=5$ & 13.1 & 3.1    & 2.2       & 2.2& 2.2    \\
$S=4$ & 6.2 & 1.6    & 1.2       & 1.2& 1.2    \\
$S=3$ & - & 0.5    & 0.5       & 0.4& 0.4    \\
$S=2$ & - & 0.9    & -0.1      & -0.1& -0.1   \\
$S=1$ & - & 0.0    & -0.1      & -0.1& -0.1   \\
$S=0$ & 0.0 & 0.0    & 0.0       & 0.0& 0.0   \\\hline
States & n/a&\multicolumn{2}{c|}{$\approx 4.1\times10^7$}&6910&3014\\
\end{tabular}   
    \caption{Spin ladder of Fe$_3$ system with (16e,15o) active space. $S=0$ state is set as a reference. Results of LAS-CASCI and LASSI[1,5] compared with previous CASSCF and CASPT2 results from Ref. \citenum{vitillo2019quantum} with the same (16e,15o) active space. Empty CASPT2 rows are indicative of results not available in Ref. \citenum{vitillo2019quantum}. Absolute and relative energies LASSI[1,5] and LASSI[1,5$_\text{CT}$] differ by less than 0.01kJ/mol. LASSI[1,5$_\text{CT}$] has 3014 states. $^\dagger$: Results from Ref. \citenum{vitillo2019quantum} use relativistic all-electron ANO-RCC triple-$\zeta$ basis sets for all O and Fe atoms, and double-$\zeta$ for C and H atoms. }

    \label{tab:Fe3_15_16}
\end{table}

Table \ref{tab:Fe3_15_16} shows the LASSI results for the (16e,15o) active space against the LAS-CASCI limit. The absolute values are reported in SI table XI. LASSI[0,1] incorrectly predicts a high-spin ground state, while the LAS-CASCI predicts a triply-degenerate low-spin ground state (i.e., singlet, triplet, and quintet). The LASSI[1,5] (6910 model states) results differ from the LAS-CASCI ($4.1\times 10^7$ determinants) results only by 0.1 kJ/mol while using four orders of magnitude fewer states. LASSI[1,5] and LASSI[1,5$_\text{CT}$] (3014 model states) differ by less than 0.01kJ/mol. Additionally, CASPT2\cite{ghigo2004modified} predicts an energy difference between the $S=0$ and $S=7$ states of 23 kJ/mol, indicating the importance of dynamical correlation. 

\begin{figure}
    \begin{tabular}{c|c|c}
        Charge & $\Delta S$ & $\bar{n}_{\mathcal{P}K}$\\\hline
        \raisebox{-.5\height}{\includegraphics[width=20mm]{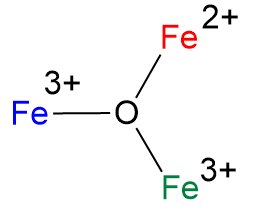}} & $\begin{bmatrix} \textcolor{red}{S^{(1)}} \\ \textcolor{blue}{S^{(2)}} \\ \textcolor{cadmiumgreen}{S^{(3)}} \end{bmatrix}=\begin{bmatrix} 2 \\ 2.5 \\ 2.5 
        \end{bmatrix}$ & $\begin{bmatrix} \textcolor{red}{0.00} \\ \textcolor{blue}{0.00} \\ \textcolor{cadmiumgreen}{0.00}  \end{bmatrix}$\\
        \raisebox{-.5\height}{\includegraphics[width=20mm]{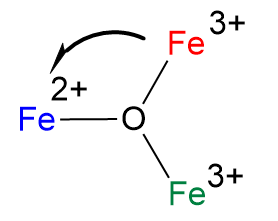}} & $\begin{bmatrix} \textcolor{red}{-0.5} \\ \textcolor{blue}{-0.5} \\ \textcolor{cadmiumgreen}{0} \end{bmatrix}^*$ &$\begin{bmatrix} \textcolor{red}{1.04} \\ \textcolor{blue}{0.91} \\ \textcolor{cadmiumgreen}{0.00}  \end{bmatrix}$\\
        \raisebox{-.5\height}{\includegraphics[width=20mm]{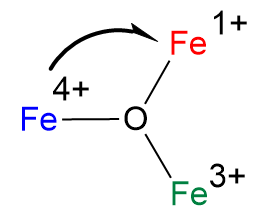}}  & $\begin{bmatrix} \textcolor{red}{-0.5} \\ \textcolor{blue}{-0.5} \\ \textcolor{cadmiumgreen}{0} \end{bmatrix}$ &$\begin{bmatrix} \textcolor{red}{1.18} \\ \textcolor{blue}{2.02} \\ \textcolor{cadmiumgreen}{0.00}  \end{bmatrix}$\\
        \raisebox{-.5\height}{\includegraphics[width=20mm]{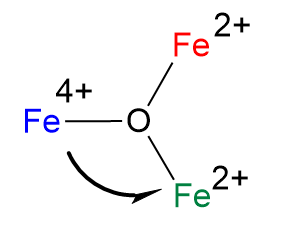}} & $\begin{bmatrix} \textcolor{red}{0} \\ \textcolor{blue}{-0.5} \\ \textcolor{cadmiumgreen}{-0.5} \end{bmatrix}$ &$\begin{bmatrix} \textcolor{red}{0.00} \\ \textcolor{blue}{1.19} \\ \textcolor{cadmiumgreen}{1.40}  \end{bmatrix}$\\
    \end{tabular}
    \caption{Summary table of LASSI[1,5](16e,15o) results for the Fe$_3$ system in different rootspaces. The first column denotes a pictorial representation of the rootspace, in which the charge on each iron is that after the electron is transferred in the direction of the arrow. The change in spin on each iron in the rootspace is shown in the 2nd column. The third column shows the average excitation number [$\bar{n}_{K\mathcal{P}}$, Eq.\ (\ref{eq:navg})] on each fragment in the given rootspace and it's all possible spin polarizations.}
    \label{fig:navg_rootspaces}

\end{figure}

However, the LASSI[1,$\nmax$] approach appears to create significant deadwood in these low-energy wave functions. We find that 48 of the 182 rootspaces generated at the $r=1$ level have zero average weight [$w_\mathcal{P}$, Eq.\ (\ref{eq:sa_weights})]; these correspond to rootspaces where the charge transfer Fe$^{2+}$ $\rightarrow$ Fe$^{3+}$ results in increased $S$ on Fe$^{2+}$ center. Additionally, we find that the average excitation number [$\bar{n}_{K\mathcal{P}}$, Eq.\ (\ref{eq:navg})] is zero if the corresponding fragment does not participate in a charge transfer compared to the reference state (Figure \ref{fig:navg_rootspaces}). In other words, for any $r=1$ rootspace (cf. Table \ref{tab:rootspaces}), a ``spectator'' fragment does not need local excitations. This observation motivates the development of LASSI[$r$,$\nmaxCT$], as described in Sec.\ \ref{sec:nmaxCT}. The result for LASSI[1,5$_{\text{CT}}$] is quantitatively the same as LASSI[1,5], but with fewer than half the model states of LASSI[1,5].

\begin{figure}
    \centering
    \includegraphics[width=\linewidth]{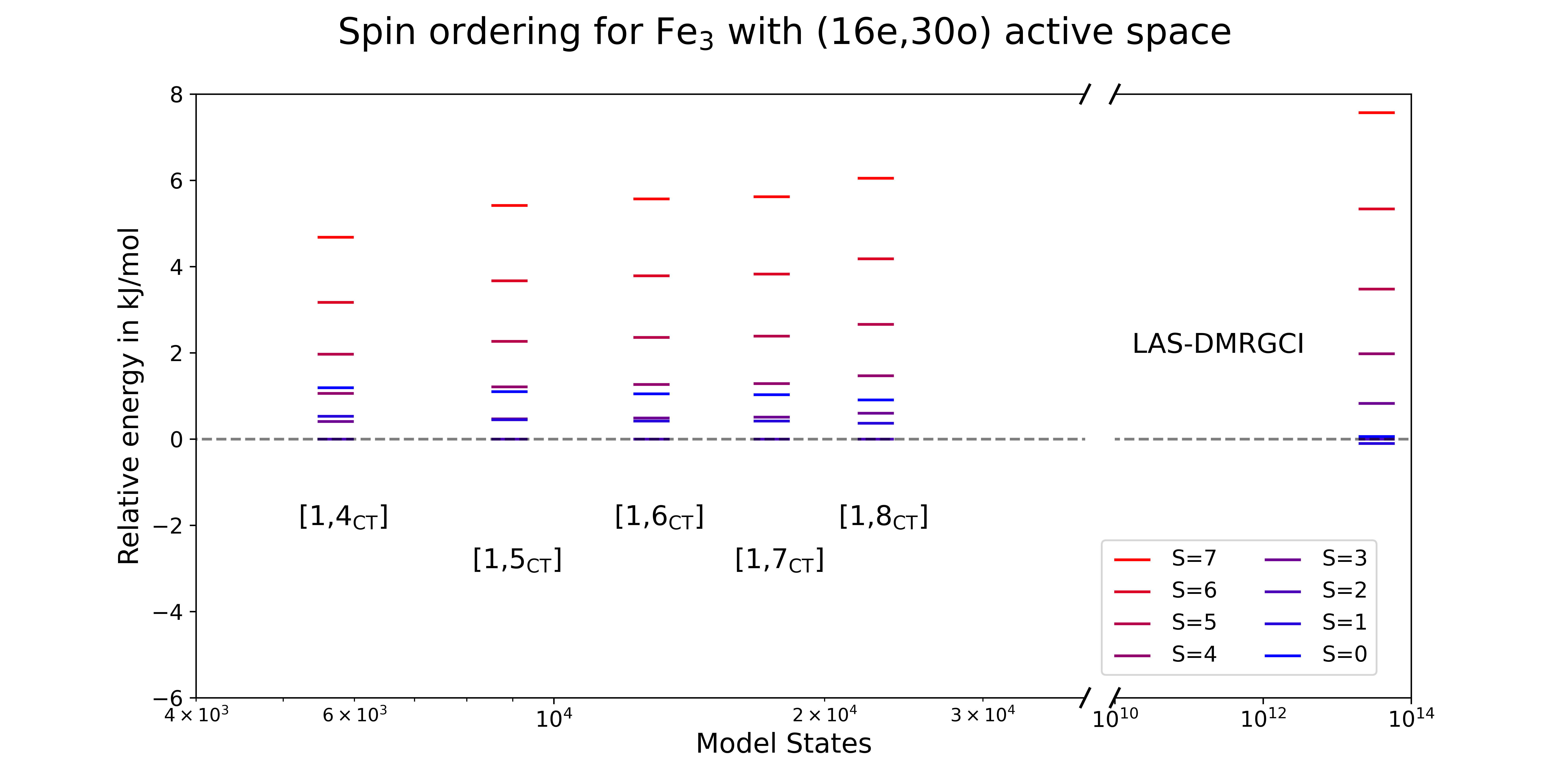}
    \caption{Spin ladder for Fe$_3$ system with (16e,30o) active space. [1,$\nmaxCT$] are the corresponding LASSI calculations. The state $S=2$ is set as a reference. DMRGCI is placed on the abscissa at the number of single determinants in the corresponding hypothetical LAS-CASCI calculation. LASSI[1,8$_\text{CT}$] recovers LAS-DMRGCI spin ladder qualitatively.
    }
    \label{fig:Fe3_30_16}
\end{figure}

We now discuss results with (16e,30o) active space presented  in Fig. \ref{fig:Fe3_30_16}. Since one cannot perform CASSCF or CASCI in this active space, we have used DMRG \cite{zhai2023block2} to calculate the reference value starting from optimized orbitals of LASSCF(16e,30o) (SI table XII). DMRG was performed with a bond dimension $M=500$. We show DMRGCI results in Fig. \ref{fig:Fe3_30_16} on the abscissa at 34 trillion states corresponding to the size of a (16e,30o) CAS calculations. We see that the qualitative features of the DMRG(16e,30o) are generally reproduced at the LASSI[1,8$_\text{CT}$] level (22808 model states), except that LASSI predicts a quintet ground state and DMRG predicts a degenerate singlet, triplet, and quintet. 
\\
\section{Conclusions and Outlook\label{sec:conclusion}}
We have introduced LASSI[$r$,$\nmax$], a systematic hierarchy of post-LASSCF methods in which the CASCI limit is recovered in the limit of large $r$ and $\nmax$. Test calculations on the spin-state energy ladder of a triiron compound Fe$_3$ and its aluminum-substituted variant AlFe$_2$ show that LASSI[$r$,$\nmax$] qualitatively reproduces CASCI for $r=1$ and $\nmax$ no greater than 15, and usually as low as 7, leading to a compactification of the wave function by several orders of magnitude.

Close examination of the LASSI[1,5](16e,15o) wave functions of Fe$_3$ led to the identification of another large category of deadwood model states, which are systematically removed in the more approximate LASSI[$r$,$\nmaxCT$] hierarchy of methods. Using this latter model, we report for the first time practical LASSI calculations on a trimetallic species with both 3$d$ and 4$d$ orbitals in the model space, a calculation which is impossible for conventional CASCI but which gives qualitatively similar results to the corresponding (16e,30o) DMRG-CI calculation. Additionally, LASSI allows us to understand the wave function and draw meaningful conclusions, which is difficult with DMRGCI.

The convergence with respect to $\nmax$ or $\nmaxCT$ appears to be sensitive to the size of the active space, in both systems. This is especially pronounced for the AlFe$_2$ system. However, the fact that there is essentially no difference in the values of the von Neumann entropies of the LASSI[1,$\nmax$] wave functions between the smaller and larger active spaces suggests that the fragments are not more strongly entangled in the latter, in a physical sense, despite the larger $\nmax$ required to achieve agreement with CASCI. Generating more compact LASSI wave functions is a primary target for future work.

\section{Acknowledgment}
\label{sec:acknowledgment}
We would like to thank Dr. Rishu Khurana, Dr. Cong Liu and Dr. Christopher Knight for helpful discussions. This work is supported as part of the Computational Chemical Sciences Program, under Award DE-SC0023382, funded by the U.S. Department of Energy, Office of Basic Energy Sciences, Chemical Sciences, Geosciences, and Biosciences Division.
\section{Supplemental Information\label{sec:SI}}
The supplemental information contains geometries of compounds studied, all converged active space orbitals for all active spaces for all compounds and absolute energies of LASSI, LAS-CASCI and LAS-DMRGCI calculations performed in this work. 
\section{Conflict of interest}
The authors declare no competing conflict of interest.
\section{Data availability}
Input and output files, converged orbitals and geometries are available on Ref. \citenum{data}.
\bibliography{main}
\begin{tocentry}
    \includegraphics[width=82.5mm, height=42.5mm]{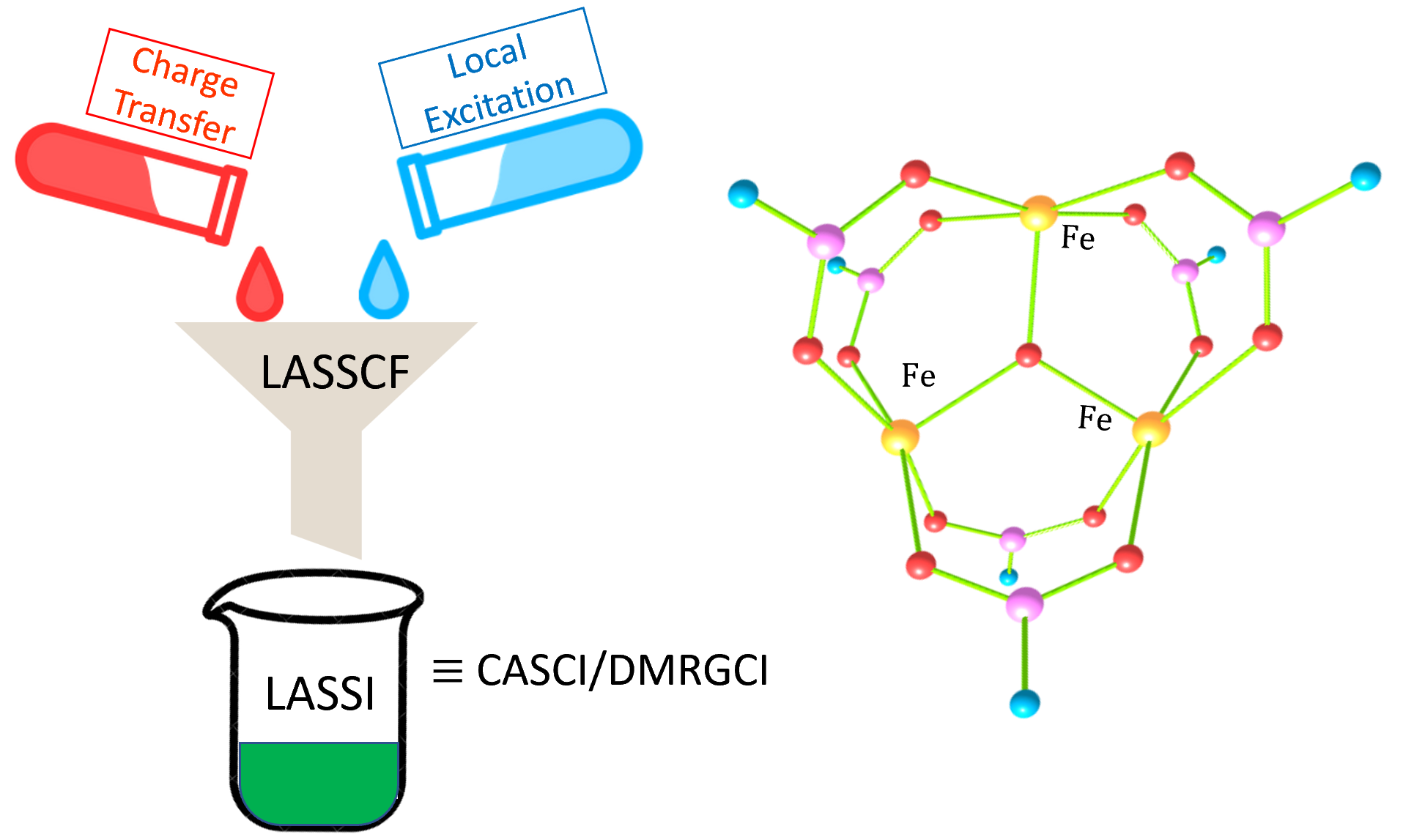}
\end{tocentry}


\section*{Supplementary material}

# Supplemental Information to: Automatic State Interaction with Large Localized Active Spaces for Multimetallic Systems

Valay Agarawal, Daniel King, Matthew R. Hermes,* and Laura Gagliardi*

*Department of Chemistry, University of Chicago.*

E-mail: mrhermes@uchicago.edu; lgagliardi@uchicago.edu

1

# Aluminium Di Iron complex

This complex has been analyzed with 3 different active spaces, (9e,9o), (11e,10o) and (11e,20o). The basis set is cc-pvdz for C and H atoms, and cc-pvtz for Al,Fe and O atoms. We provide the following tables in this section.

1. Geometry of molecule.

2. Absolute energies of the all spin states and $J$ coupling with various LASSI calculations in Table II of main manuscript.

3. Absolute energies of all spin states and $J$ coupling of various LASSI and LAS-CASCI calculations in Fig 2 of main manuscript.

4. Optimized LASSCF active space orbitals and their occupations pertaining to (11e,10o) active space.

5. Optimized LASSCF active space orbitals and their occupations pertaining to (9e,9o) active space.

6. Optimized LASSCF active space orbitals and their occupations pertaining to (11e,20o) active space.

7. Entropy and $q^{(\text{avg})}$ for all rootspaces pertaining to (11e,10o) with LASSI[$r=1, q=5$]

8. Entropy and $q^{(\text{avg})}$ for all rootspaces pertaining to (11e,20o) with LASSI[$r=1, q=15$]

Table 1: Geometry of Al-Fe-Fe MOF Node. All values in Å

| | | | |
|---|---|---|---|
| O  | -2.220198244  | 0.3991903003  | 1.694471699 |
| O  | -1.685553245  | -1.782306322  | 1.431399507 |
| C  | -2.268517865  | -0.8319550379 | 1.983951274 |
| H  | -2.913342017  | -1.076728589  | 2.843786805 |
| O  | 1.388288081   | 2.079556156   | -1.347085675 |
| O  | -0.7599088595 | 2.580923635   | -0.849227704 |
| C  | 0.3465674686  | 2.753895325   | -1.438835699 |
| H  | 0.3723796145  | 3.61769966    | -2.12309115 |
| O  | 1.046960908   | -2.6425342    | 0.9613246722 |
| O  | -0.3991903003 | 2.220198244   | 1.694471699 |
| O  | 1.782306322   | 1.685553245   | 1.431399507 |
| C  | 0.8319550379  | 2.268517865   | 1.983951274 |
| H  | 1.076728589   | 2.913342017   | 2.843786805 |
| O  | -2.079556156  | -1.388288081  | -1.347085675 |
| O  | -2.580923635  | 0.7599088595  | -0.849227704 |
| C  | -2.753895325  | -0.3465674686 | -1.438835699 |
| H  | -3.61769966   | -0.3723796145 | -2.12309115 |
| Fe | -0.3798741893 | -1.792603363  | -0.2003377303 |
| O  | 0.6422231803  | -2.237796553  | -1.892914518 |
| Fe | 1.792603363   | 0.3798741893  | -0.2003377303 |
| O  | 2.237796553   | -0.6422231803 | -1.892914518 |
| O  | 2.6425342     | -1.046960908  | 0.9613246722 |
| Al | -1.236271642  | 1.236271642   | 0.350650148 |
| O  | -0.0449501954 | 0.0449501954  | 0.0127621413 |
| C  | 1.645528685   | -1.645528685  | -2.357112465 |
| H  | 2.056906298   | -2.056906298  | -3.294571292 |
| C  | 2.160924281   | -2.160924281  | 1.277584187 |
| H  | 2.796080543   | -2.796080543  | 1.918244302 |

Table 2: Total Energy + 3971 $E_h$ for all states for various LASSI calculations with $r$ and $q$ and $J$ value with AlFe$_2$ with (11e,10o) active space.

| r | n | States | S=9/2 | S=7/2 | S=5/2 | S=3/2 | S=1/2 | J |
|---|---|---|---|---|---|---|---|---|
| r=0 | q=1 | 5 | -0.792354 | -0.787569 | -0.783847 | -0.781189 | -0.779594 | 116.69 |
| | q=2 | 10 | -0.792354 | -0.787569 | -0.783847 | -0.781189 | -0.779594 | 116.69 |
| | q=3 | 15 | -0.792354 | -0.787569 | -0.783847 | -0.781189 | -0.779594 | 116.69 |
| | q=4 | 20 | -0.792354 | -0.787569 | -0.783847 | -0.781189 | -0.779594 | 116.69 |
| | q=5 | 25 | -0.792354 | -0.787569 | -0.783847 | -0.781189 | -0.779594 | 116.69 |
| | q=6 | 25 | -0.792354 | -0.787569 | -0.783847 | -0.781189 | -0.779594 | 116.69 |
| | q=7 | 25 | -0.792354 | -0.787569 | -0.783847 | -0.781189 | -0.779594 | 116.69 |
| | q=8 | 25 | -0.792354 | -0.787569 | -0.783847 | -0.781189 | -0.779594 | 116.69 |
| | q=9 | 25 | -0.792354 | -0.787569 | -0.783847 | -0.781189 | -0.779594 | 116.69 |
| | q=10 | 25 | -0.792354 | -0.787569 | -0.783847 | -0.781189 | -0.779594 | 116.69 |
| r=1 | q=1 | 18 | -0.792354 | -0.787674 | -0.783994 | -0.781369 | -0.779817 | 114.65 |
| | q=2 | 52 | -0.792354 | -0.788224 | -0.785009 | -0.782712 | -0.781336 | 100.75 |
| | q=3 | 102 | -0.792354 | -0.790834 | -0.789668 | -0.788841 | -0.788318 | 36.91 |
| | q=4 | 168 | -0.792354 | -0.792213 | -0.792078 | -0.791969 | -0.791898 | 4.17 |
| | q=5 | 250 | -0.792354 | -0.792507 | -0.792598 | -0.792647 | -0.792671 | -2.90 |
| | q=6 | 290 | -0.792354 | -0.792507 | -0.792598 | -0.792647 | -0.792671 | -2.90 |
| | q=7 | 330 | -0.792354 | -0.792507 | -0.792598 | -0.792647 | -0.792671 | -2.90 |
| | q=8 | 370 | -0.792354 | -0.792507 | -0.792598 | -0.792647 | -0.792671 | -2.90 |
| | q=9 | 410 | -0.792354 | -0.792507 | -0.792598 | -0.792647 | -0.792671 | -2.90 |
| | q=10 | 450 | -0.792354 | -0.792507 | -0.792598 | -0.792647 | -0.792671 | -2.90 |
| r=2 | q=1 | 38 | -0.792354 | -0.787676 | -0.783996 | -0.78137 | -0.779817 | 114.65 |
| | q=2 | 126 | -0.792354 | -0.788227 | -0.785013 | -0.782715 | -0.781339 | 100.73 |
| | q=3 | 264 | -0.792354 | -0.790836 | -0.789673 | -0.78885 | -0.788329 | 36.81 |
| | q=4 | 452 | -0.792354 | -0.792214 | -0.792089 | -0.791994 | -0.791932 | 3.85 |
| | q=5 | 690 | -0.792354 | -0.792509 | -0.792612 | -0.792677 | -0.792712 | -3.28 |
| | q=6 | 872 | -0.792354 | -0.79251 | -0.792618 | -0.792689 | -0.79273 | -3.43 |
| | q=7 | 1072 | -0.792354 | -0.79251 | -0.792621 | -0.792696 | -0.792739 | -3.52 |
| | q=8 | 1290 | -0.792354 | -0.79251 | -0.792623 | -0.792701 | -0.792746 | -3.58 |
| | q=9 | 1526 | -0.792354 | -0.79251 | -0.792626 | -0.792707 | -0.792754 | -3.66 |
| | q=10 | 1780 | -0.792354 | -0.79251 | -0.792628 | -0.792711 | -0.79276 | -3.71 |
| r=3 | q=1 | 62 | -0.792354 | -0.787676 | -0.783996 | -0.78137 | -0.779817 | 114.65 |
| | q=2 | 216 | -0.792354 | -0.788227 | -0.785013 | -0.782715 | -0.781339 | 100.73 |
| | q=3 | 462 | -0.792354 | -0.790836 | -0.789673 | -0.78885 | -0.788329 | 36.81 |
| | q=4 | 800 | -0.792354 | -0.792214 | -0.792089 | -0.791994 | -0.791932 | 3.85 |
| | q=5 | 1230 | -0.792354 | -0.792509 | -0.792612 | -0.792677 | -0.792712 | -3.28 |
| | q=6 | 1610 | -0.792354 | -0.79251 | -0.792618 | -0.792689 | -0.79273 | -3.43 |
| | q=7 | 2038 | -0.792354 | -0.79251 | -0.792621 | -0.792696 | -0.792739 | -3.52 |
| | q=8 | 2514 | -0.792354 | -0.79251 | -0.792623 | -0.792701 | -0.792746 | -3.58 |
| | q=9 | 3038 | -0.792354 | -0.79251 | -0.792626 | -0.792707 | -0.792754 | -3.66 |
| | q=10 | 3610 | -0.792354 | -0.79251 | -0.792628 | -0.792711 | -0.79276 | -3.71 |
| CASCI | | 52920 | -0.792354 | -0.792511 | -0.792632 | | -0.792771 | -3.81 |

Table 3: Total Energy + 3971 $E_h$ for all states for various LASSI calculations with $r$ and $q$ and $J$ value with AlFe$_2$ with (11e,10o) active space.

| r | n | States | S=9/2 | S=7/2 | S=5/2 | S=3/2 | S=1/2 | J |
|---|---|---|---|---|---|---|---|---|
| r=0 | q=1 | 5 | -0.792354 | -0.787569 | -0.783847 | -0.781189 | -0.779594 | 116.69 |
| | q=2 | 10 | -0.792354 | -0.787569 | -0.783847 | -0.781189 | -0.779594 | 116.69 |
| | q=3 | 15 | -0.792354 | -0.787569 | -0.783847 | -0.781189 | -0.779594 | 116.69 |
| | q=4 | 20 | -0.792354 | -0.787569 | -0.783847 | -0.781189 | -0.779594 | 116.69 |
| | q=5 | 25 | -0.792354 | -0.787569 | -0.783847 | -0.781189 | -0.779594 | 116.69 |
| | q=6 | 25 | -0.792354 | -0.787569 | -0.783847 | -0.781189 | -0.779594 | 116.69 |
| | q=7 | 25 | -0.792354 | -0.787569 | -0.783847 | -0.781189 | -0.779594 | 116.69 |
| | q=8 | 25 | -0.792354 | -0.787569 | -0.783847 | -0.781189 | -0.779594 | 116.69 |
| | q=9 | 25 | -0.792354 | -0.787569 | -0.783847 | -0.781189 | -0.779594 | 116.69 |
| | q=10 | 25 | -0.792354 | -0.787569 | -0.783847 | -0.781189 | -0.779594 | 116.69 |
| r=1 | q=1 | 18 | -0.792354 | -0.787674 | -0.783994 | -0.781369 | -0.779817 | 114.65 |
| | q=2 | 52 | -0.792354 | -0.788224 | -0.785009 | -0.782712 | -0.781336 | 100.75 |
| | q=3 | 102 | -0.792354 | -0.790834 | -0.789668 | -0.788841 | -0.788318 | 36.91 |
| | q=4 | 168 | -0.792354 | -0.792213 | -0.792078 | -0.791969 | -0.791898 | 4.17 |
| | q=5 | 250 | -0.792354 | -0.792507 | -0.792598 | -0.792647 | -0.792671 | -2.90 |
| | q=6 | 290 | -0.792354 | -0.792507 | -0.792598 | -0.792647 | -0.792671 | -2.90 |
| | q=7 | 330 | -0.792354 | -0.792507 | -0.792598 | -0.792647 | -0.792671 | -2.90 |
| | q=8 | 370 | -0.792354 | -0.792507 | -0.792598 | -0.792647 | -0.792671 | -2.90 |
| | q=9 | 410 | -0.792354 | -0.792507 | -0.792598 | -0.792647 | -0.792671 | -2.90 |
| | q=10 | 450 | -0.792354 | -0.792507 | -0.792598 | -0.792647 | -0.792671 | -2.90 |
| r=2 | q=1 | 38 | -0.792354 | -0.787676 | -0.783996 | -0.78137 | -0.779817 | 114.65 |
| | q=2 | 126 | -0.792354 | -0.788227 | -0.785013 | -0.782715 | -0.781339 | 100.73 |
| | q=3 | 264 | -0.792354 | -0.790836 | -0.789673 | -0.78885 | -0.788329 | 36.81 |
| | q=4 | 452 | -0.792354 | -0.792214 | -0.792089 | -0.791994 | -0.791932 | 3.85 |
| | q=5 | 690 | -0.792354 | -0.792509 | -0.792612 | -0.792677 | -0.792712 | -3.28 |
| | q=6 | 872 | -0.792354 | -0.79251 | -0.792618 | -0.792689 | -0.79273 | -3.43 |
| | q=7 | 1072 | -0.792354 | -0.79251 | -0.792621 | -0.792696 | -0.792739 | -3.52 |
| | q=8 | 1290 | -0.792354 | -0.79251 | -0.792623 | -0.792701 | -0.792746 | -3.58 |
| | q=9 | 1526 | -0.792354 | -0.79251 | -0.792626 | -0.792707 | -0.792754 | -3.66 |
| | q=10 | 1780 | -0.792354 | -0.79251 | -0.792628 | -0.792711 | -0.79276 | -3.71 |
| r=3 | q=1 | 62 | -0.792354 | -0.787676 | -0.783996 | -0.78137 | -0.779817 | 114.65 |
| | q=2 | 216 | -0.792354 | -0.788227 | -0.785013 | -0.782715 | -0.781339 | 100.73 |
| | q=3 | 462 | -0.792354 | -0.790836 | -0.789673 | -0.78885 | -0.788329 | 36.81 |
| | q=4 | 800 | -0.792354 | -0.792214 | -0.792089 | -0.791994 | -0.791932 | 3.85 |
| | q=5 | 1230 | -0.792354 | -0.792509 | -0.792612 | -0.792677 | -0.792712 | -3.28 |
| | q=6 | 1610 | -0.792354 | -0.79251 | -0.792618 | -0.792689 | -0.79273 | -3.43 |
| | q=7 | 2038 | -0.792354 | -0.79251 | -0.792621 | -0.792696 | -0.792739 | -3.52 |
| | q=8 | 2514 | -0.792354 | -0.79251 | -0.792623 | -0.792701 | -0.792746 | -3.58 |
| | q=9 | 3038 | -0.792354 | -0.79251 | -0.792626 | -0.792707 | -0.792754 | -3.66 |
| | q=10 | 3610 | -0.792354 | -0.79251 | -0.792628 | -0.792711 | -0.79276 | -3.71 |
| CASCI | | 52920 | -0.792354 | -0.792511 | -0.792632 | | -0.792771 | -3.81 |

Table 4: LASSCF Optimized Active Space orbitals and their occupancies with the active space of (11e,10o). Isosurface value = 0.03. Note the shape of doubly occupied orbital. It is poorly defined, optimizing a fully filled or completely empty orbital is challenging since the energy is invariant to the rotation of those orbitals.

| | | | |
|---|---|---|---|
| 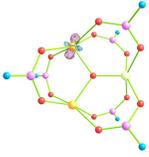 2.0000 | 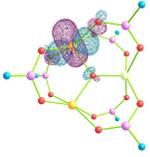 1.0000 | 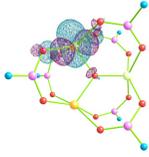 1.0000 | 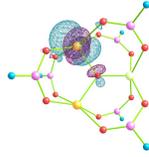 1.0000 |
| 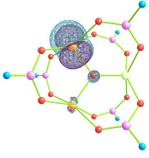 1.0000 | 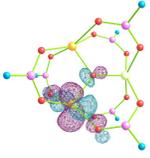 1.0000 | 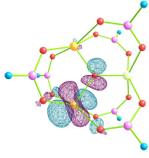 1.0000 | 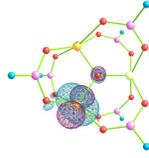 1.0000 |
| | 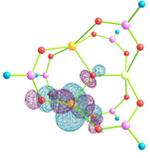 1.0000 | 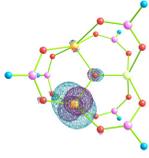 1.0000 | |

Table 5: LASSCF Optimized Active Space orbitals and their occupancies with the active space of (9e,9o). Isosurface value = 0.03

| | | | |
|---|---|---|---|
| 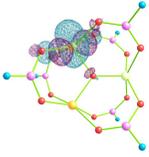 1.0000 | 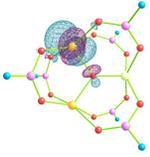 1.0000 | 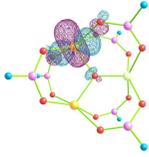 1.0000 | 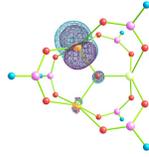 1.0000 |
| 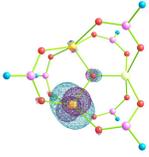 1.0000 | 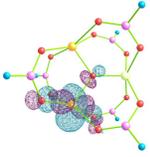 1.0000 | 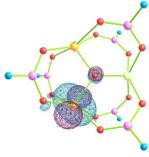 1.0000 | 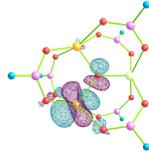 1.0000 |
| | 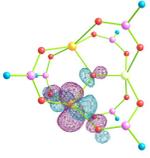 1.0000 | | |

Table 6: LASSCF Optimized Active Space orbitals and their occupancies with the active space of (11e,20o). Isosurface value = 0.03

| | | | |
|---|---|---|---|
| 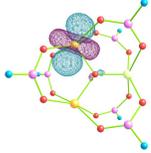 | 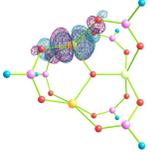 | 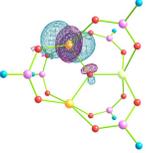 | 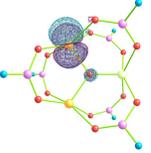 |
| 1.98777 | 0.99658 | 0.99621 | 0.99601 |
| 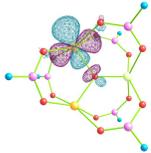 | 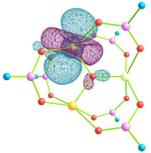 | 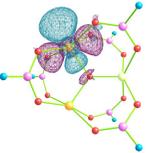 | 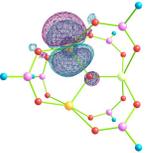 |
| 0.99583 | 0.01195 | 0.00423 | 0.00406 |
| 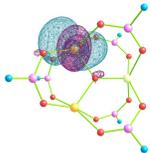 | 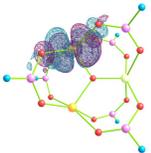 | 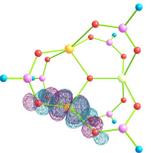 | 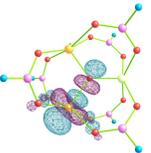 |
| 0.00389 | 0.00347 | 0.99789 | 0.99777 |
| 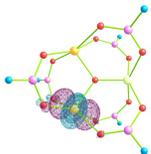 | 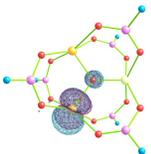 | 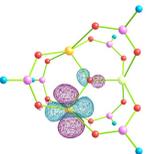 | 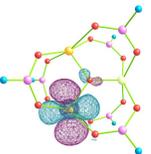 |
| 0.99774 | 0.99772 | 0.99772 | 0.00228 |
| 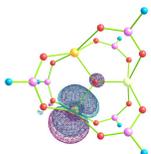 | 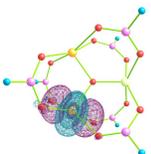 | 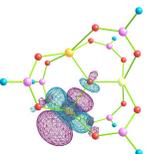 | 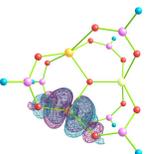 |
| 0.00228 | 0.00226 | 0.00223 | 0.00211 |

Table 7: Entropy for AlFeFe node with (11e,10o) active space for LASSI[1,5]. $Fe^{2+}$ in red and $Fe^{3+}$ in blue. $s_{K\mathcal{P}}$ is entropy for $K$th fragment in $\mathcal{P}$ rootspace. $q^{(\text{avg})}$ is also provided for reference. 5 rootspaces are not shown due to low weights.

| Rootspace | $s_{\mathcal{P}K}$ | | $\bar{q}_{\mathcal{P}K}$ | |
|---|---|---|---|---|
| $Fe^{3+}$—O—$Fe^{2+}$ / Al$^{3+}$ | 0.00 | 0.00 | 0.00 | 0.00 |
| | 0.00 | 0.00 | 0.00 | 0.00 |
| | 0.00 | 0.00 | 0.00 | 0.00 |
| | 0.00 | 0.00 | 0.00 | 0.00 |
| | 0.00 | 0.00 | 0.00 | 0.00 |
| $Fe^{4+}$—O—$Fe^{1+}$ / Al$^{3+}$ | 0.77 | 0.77 | 0.87 | 1.87 |
| | 0.79 | 0.79 | 0.95 | 1.87 |
| | 0.79 | 0.79 | 0.99 | 1.87 |
| | 0.79 | 0.79 | 0.91 | 1.87 |
| $Fe^{2+}$—O—$Fe^{3+}$ / Al$^{3+}$ | 1.03 | 1.03 | 1.20 | 1.11 |
| | 1.02 | 1.02 | 1.21 | 1.11 |
| | 1.02 | 1.02 | 1.21 | 1.12 |
| | 1.02 | 1.02 | 1.21 | 1.10 |

Table 8: Entropy for AlFeFe node with (11e,20o) active space with LASSI[1,15]. $Fe^{2+}$ in red and $Fe^{3+}$ in blue. $s_{K\mathcal{P}}$ is entropy for $K$th fragment in $\mathcal{P}$ rootspace. $q^{(\mathrm{avg})}$ is also provided for reference. Note that even though $n^{(\mathrm{avg})}$ can be really high, the entropy is close to 1. updated

| Rootspace | $s_{\mathcal{P}K}$ | | $\bar{q}_{\mathcal{P}K}$ | |
|---|---|---|---|---|
| 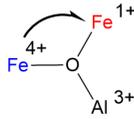 | 0.14 | 0.14 | 2.27 | 1.68 |
| | 0.82 | 0.82 | 2.09 | 2.14 |
| | 0.14 | 0.14 | 2.27 | 1.70 |
| | 0.82 | 0.82 | 2.10 | 2.14 |
| | 0.14 | 0.14 | 2.27 | 1.73 |
| | 0.82 | 0.82 | 2.10 | 2.15 |
| | 0.22 | 0.16 | 2.30 | 1.70 |
| | 0.14 | 0.14 | 2.27 | 1.72 |
| | 0.82 | 0.82 | 2.10 | 2.14 |
| 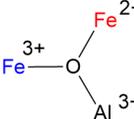 | 0.00 | 0.00 | 0.00 | 0.00 |
| | 0.00 | 0.00 | 0.00 | 0.00 |
| | 0.00 | 0.00 | 0.00 | 0.00 |
| | 0.00 | 0.00 | 0.00 | 0.00 |
| | 0.00 | 0.00 | 0.00 | 0.00 |
| 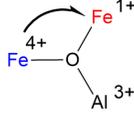 | 0.13 | 0.13 | 0.16 | 2.69 |
| | 0.13 | 0.13 | 0.16 | 2.69 |
| | 0.22 | 0.34 | 6.98 | 1.86 |
| | 0.00 | 0.00 | 0.02 | 1.83 |
| | 0.13 | 0.13 | 0.16 | 2.69 |
| | 1.01 | 0.92 | 5.66 | 1.20 |
| | 0.20 | 0.21 | 7.01 | 2.09 |
| | 0.00 | 0.06 | 0.02 | 1.72 |
| | 0.13 | 0.13 | 0.16 | 2.69 |
| | 0.00 | 0.00 | 0.02 | 1.75 |
| | 0.22 | 0.25 | 6.98 | 2.24 |
| | 1.01 | 0.92 | 5.69 | 1.20 |
| | 0.13 | 0.13 | 0.16 | 2.69 |
| | 0.00 | 0.00 | 0.02 | 1.80 |
| | 0.24 | 0.37 | 6.88 | 2.28 |
| | 0.98 | 0.91 | 5.57 | 1.18 |
| | 0.13 | 0.13 | 0.16 | 2.69 |
| | 1.06 | 0.93 | 5.74 | 1.21 |
| | 0.13 | 0.13 | 0.16 | 2.69 |

# Fe-Fe-Fe complex

This complex has been analyzed with 2 different active spaces, (16e,15o) and (16e,30o). The basis set is cc-pvdz for C and H atoms, and cc-pvtz for Al,Fe and O atoms. We provide 5 tables in this section.

1. Geometry of molecule.

2. Optimized LASSCF active space orbitals and their occupations pertaining to (16e,15o) active space.

3. Absolute energies of LASSI/CASCI calculations and states pertaining to them for (16e,15o) active space

4. Optimized LASSCF active space orbitals and their occupations pertaining to (16e,30o) active space.

5. Absolute energies of calculations pertaining to (16e,30o) active space.

6. $q^{(\text{avg})}$ for all $\mathcal{P}$ with considerable weights in LASSI[1,5] for Fe$_3$ with active space of (16e,15o).

Table 9: Geometry of Fe-Fe-Fe MOF Node. All values in Å

|    |              |              |              |
|----|--------------|--------------|--------------|
| O  | -1.396530536 | -2.155053668 | 1.497463409  |
| O  | -1.424929948 | -0.1579202181| 2.556757216  |
| C  | -1.805663158 | -1.333357733 | 2.354770591  |
| H  | -2.611139824 | -1.698287377 | 3.015172572  |
| O  | 1.422169433  | -0.1727766066| -2.448307975 |
| O  | 1.404243292  | -2.165442677 | -1.387375376 |
| C  | 1.812327533  | -1.345288486 | -2.246529164 |
| H  | 2.621336914  | -1.706173551 | -2.904090708 |
| O  | -1.428899418 | 2.343625706  | 1.186473538  |
| O  | -1.403562858 | -2.165307125 | -1.387572278 |
| O  | -1.422998602 | -0.1723629226| -2.447964967 |
| C  | -1.812500488 | -1.345105901 | -2.246279414 |
| H  | -2.621699381 | -1.706195274 | -2.903496624 |
| O  | 1.425112371  | -0.1571948062| 2.556693752  |
| O  | 1.397382095  | -2.154305543 | 1.497360669  |
| C  | 1.806221904  | -1.332524996 | 2.35469992   |
| H  | 2.611813123  | -1.697184248 | 3.015110605  |
| Fe | -0.0001429864| 0.9716426461 | 1.624782043  |
| O  | 1.428198549  | 2.344050002  | 1.186026051  |
| Fe | -0.0001364224| 0.952927878  | -1.522872889 |
| O  | 1.423595107  | 2.330128125  | -1.070103354 |
| O  | -1.423503945 | 2.330456754  | -1.069656998 |
| Fe | 0.0002549988 | -1.941619777 | 0.0458885284 |
| O  | -0.0001521382| -0.0719149875| 0.0721478962 |
| C  | 1.815220274  | 2.721515148  | 0.0544222441 |
| H  | 2.604057215  | 3.492625269  | 0.0488314748 |
| C  | -1.815568952 | 2.721425035  | 0.0548464944 |
| H  | -2.604489027 | 3.492449072  | 0.0492507598 |

Table 10: LASSCF Optimized Active Space orbitals and their occupancies with the active space of (16e,15o). Isosurface value = 0.03

| 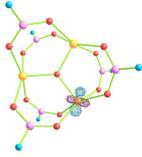 | 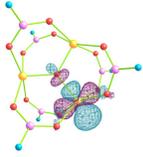 | 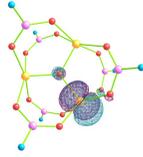 | 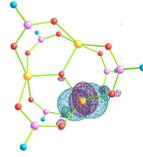 |
|---|---|---|---|
| 2.0000 | 1.0000 | 1.0000 | 1.0000 |
| 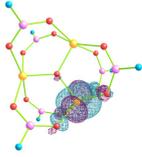 | 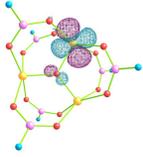 | 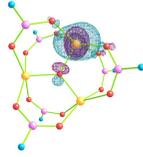 | 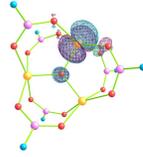 |
| 1.0000 | 1.0000 | 1.0000 | 1.0000 |
| 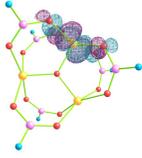 | 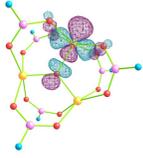 | 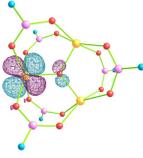 | 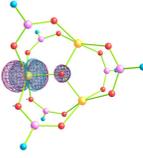 |
| 1.0000 | 1.0000 | 1.0000 | 1.0000 |
| 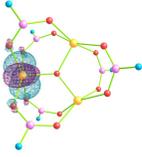 | 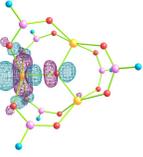 | 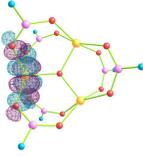 | |
| 2.0000 | 1.0000 | 1.0000 | |

Table 11: Absolute energies for LASSI and LAS-CASCI upto 6 decimal digits in $Fe_3$ system with (16e,15o) active space.

| Spin | LASSI[0,1] | LASSI[1,1] | LASSI[1,5] | LASSI[1,5$_{CT}$] | LAS-CASCI |
|---|---|---|---|---|---|
| S=7 | -4992.234449 | -4992.234449 | -4992.234449 | -4992.234449 | -4992.234449 |
| S=6 | -4992.233872 | -4992.234049 | -4992.235012 | -4992.235012 | -4992.235015 |
| S=5 | -4992.233378 | -4992.233695 | -4992.235479 | -4992.235479 | -4992.235489 |
| S=4 | -4992.232966 | -4992.233392 | -4992.235854 | -4992.235854 | -4992.235871 |
| S=3 | -4992.232637 | -4992.233143 | -4992.236140 | -4992.236140 | -4992.236165 |
| S=2 | -4992.232127 | -4992.232597 | -4992.236344 | -4992.236344 | -4992.236376 |
| S=1 | -4992.231751 | -4992.232181 | -4992.236360 | -4992.236360 | -4992.236394 |
| S=0 | -4992.231511 | -4992.231869 | -4992.236311 | -4992.236311 | -4992.236345 |
| States | 24 | 182 | 6910 | 3014 | 41.4 million |

Table 12: LASSCF Optimized Active Space orbitals and their occupancies with the active space of (16e,30o). Isosurface value = 0.03

| | | | |
|---|---|---|---|
| 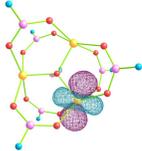 1.98795 | 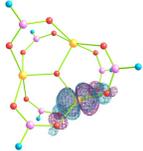 0.99659 | 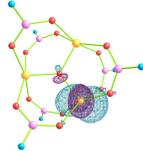 0.99620 | 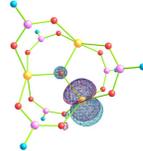 0.99605 |
| 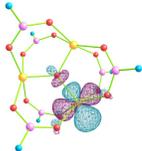 0.99581 | 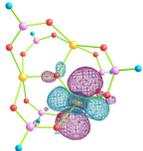 0.01177 | 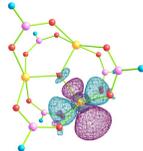 0.00423 | 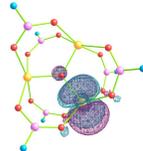 0.00403 |
| 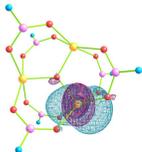 0.00390 | 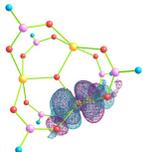 0.00346 | 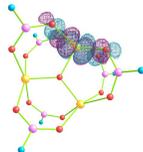 0.99790 | 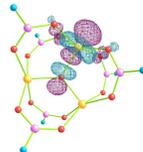 0.99777 |
| 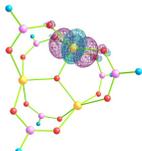 0.99774 | 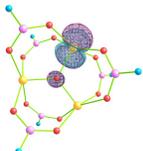 0.99772 | 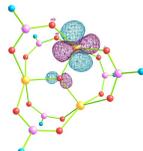 0.99772 | 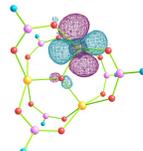 0.00228 |
| 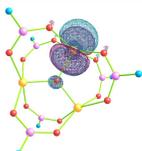 0.00228 | 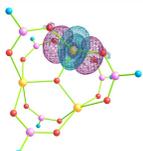 0.00226 | 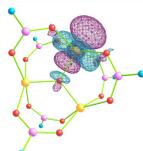 0.00223 | 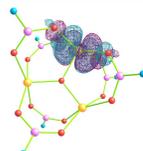 0.00210 |
| 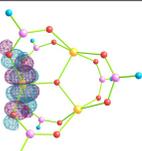 0.99790 | 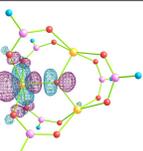 0.99775 | 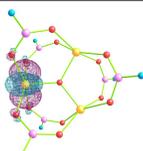 0.99773 | 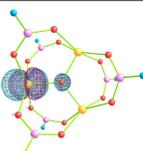 0.99771 |
| 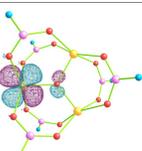 0.99771 | 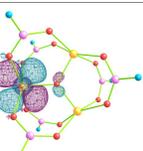 0.00229 | 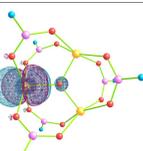 0.00229 | 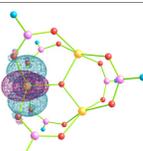 0.00227 |
| | 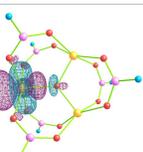 0.00225 | 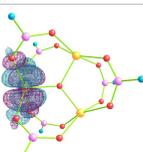 0.00210 | |

13

Table 14: (Absolute energies + 4992.36 $E_h$)×100 for LASSI and LAS-DMRGCI upto 6 decimal digits in Fe$_3$ system with (16e,30o) active space.

| | LASSI | | | | | | | | |
|---|---|---|---|---|---|---|---|---|---|
| Spin | [0,1] | [1,1] | [1,3$_{\text{CT}}$] | [1,4$_{\text{CT}}$] | [1,5$_{\text{CT}}$] | [1,6$_{\text{CT}}$] | [1,7$_{\text{CT}}$] | [1,8$_{\text{CT}}$] | DMRGCI |
| 7 | -0.5901 | -0.5942 | -0.5964 | -0.6269 | -0.6271 | -0.6287 | -0.6289 | -0.6290 | -0.6985 |
| 6 | -0.5324 | -0.5382 | -0.5782 | -0.6841 | -0.6938 | -0.6965 | -0.6970 | -0.7001 | -0.7835 |
| 5 | -0.4830 | -0.4901 | -0.5623 | -0.7300 | -0.7472 | -0.7509 | -0.7518 | -0.7580 | -0.8544 |
| 4 | -0.4418 | -0.4500 | -0.5488 | -0.7648 | -0.7877 | -0.7924 | -0.7937 | -0.8031 | -0.9114 |
| 3 | -0.4088 | -0.4178 | -0.5378 | -0.7893 | -0.8159 | -0.8219 | -0.8234 | -0.8363 | -0.9552 |
| 2 | -0.3596 | -0.3712 | -0.5295 | -0.8050 | -0.8334 | -0.8408 | -0.8427 | -0.8591 | -0.9868 |
| 1 | -0.3235 | -0.3167 | -0.5196 | -0.7846 | -0.8166 | -0.8245 | -0.8269 | -0.8453 | -0.9906 |
| 0 | -0.3006 | -0.3088 | -0.5133 | -0.7597 | -0.7917 | -0.8007 | -0.8036 | -0.8246 | -0.9846 |
| States | 24 | 380 | 3228 | 5720 | 8924 | 12840 | 17468 | 22808 | 34 Trillion |

Table 15: Average $q^{(\text{avg})}$ for all roostpaces with considerable weights for LASSI[1,5] for $Fe_3$ with active space of (16e,15o). 24 rootspaces of the with no charge tranfer had $q^{(\text{avg})}$ of less than $10^{-6}$ for all fragments. 134 out of 182 rootspaces are shown. Remaining 38 rootspaces have low weight and are omitted.

| Rootspace Type | $\bar{q}_{\mathcal{P}0}$ | $\bar{q}_{\mathcal{P}1}$ | $\bar{q}_{\mathcal{P}2}$ | Rootspace Type | $\bar{q}_{\mathcal{P}0}$ | $\bar{q}_{\mathcal{P}1}$ | $\bar{q}_{\mathcal{P}2}$ |
|---|---|---|---|---|---|---|---|
| 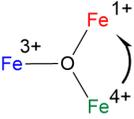 | 1.62 | 0.00 | 1.64 | 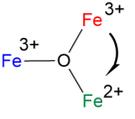 | 0.63 | 0.00 | 1.17 |
| | 1.62 | 0.00 | 1.64 | | 0.63 | 0.00 | 1.17 |
| | 1.62 | 0.00 | 1.64 | | 0.64 | 0.00 | 1.17 |
| | 1.61 | 0.00 | 1.63 | | 0.63 | 0.00 | 1.17 |
| | 1.62 | 0.00 | 1.64 | | 0.63 | 0.00 | 1.17 |
| | 1.62 | 0.00 | 1.64 | | 0.63 | 0.00 | 1.17 |
| | 1.62 | 0.00 | 1.64 | | 0.63 | 0.00 | 1.17 |
| | 1.62 | 0.00 | 1.64 | | 0.63 | 0.00 | 1.16 |
| | 1.62 | 0.00 | 1.64 | | 0.63 | 0.00 | 1.18 |
| | 1.62 | 0.00 | 1.65 | | 0.63 | 0.00 | 1.17 |
| | 1.62 | 0.00 | 1.64 | | 0.63 | 0.00 | 1.16 |
| | 1.62 | 0.00 | 1.64 | | 0.63 | 0.00 | 1.17 |
| | 1.62 | 0.00 | 1.64 | | 0.63 | 0.00 | 1.18 |
| | 1.62 | 0.00 | 1.64 | | 0.63 | 0.00 | 1.16 |
| | 1.62 | 0.00 | 1.64 | | 0.63 | 0.00 | 1.17 |
| | 1.62 | 0.00 | 1.64 | | 0.57 | 0.00 | 1.20 |
| | 1.62 | 0.00 | 1.64 | | 0.63 | 0.00 | 1.17 |
| | 1.62 | 0.00 | 1.64 | | 0.63 | 0.00 | 1.17 |
| 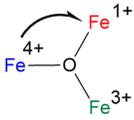 | 1.20 | 2.03 | 0.00 | 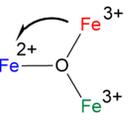 | 1.05 | 0.90 | 0.00 |
| | 1.15 | 2.01 | 0.00 | | 1.04 | 0.91 | 0.00 |
| | 1.20 | 2.03 | 0.00 | | 1.05 | 0.91 | 0.00 |
| | 1.19 | 2.03 | 0.00 | | 1.04 | 0.90 | 0.00 |
| | 1.18 | 2.02 | 0.00 | | 1.05 | 0.91 | 0.00 |
| | 1.15 | 2.01 | 0.00 | | 1.04 | 0.91 | 0.00 |
| | 1.18 | 1.96 | 0.00 | | 1.06 | 0.91 | 0.00 |
| | 1.19 | 2.03 | 0.00 | | 1.06 | 0.91 | 0.00 |
| | 1.18 | 2.02 | 0.00 | | 1.05 | 0.91 | 0.00 |
| | 1.18 | 2.02 | 0.00 | | 1.09 | 0.94 | 0.00 |
| | 1.18 | 2.02 | 0.00 | | 0.94 | 0.83 | 0.00 |
| | 1.18 | 2.02 | 0.00 | | 1.00 | 0.89 | 0.00 |
| | 1.18 | 2.02 | 0.00 | | 1.00 | 0.88 | 0.00 |
| | 1.17 | 2.01 | 0.00 | | 1.08 | 0.97 | 0.00 |
| | 1.18 | 2.02 | 0.00 | | 1.07 | 0.96 | 0.00 |
| | 1.18 | 2.02 | 0.00 | | 1.07 | 0.92 | 0.00 |
| | 1.18 | 2.02 | 0.00 | | 1.06 | 0.91 | 0.00 |
| | 1.18 | 2.02 | 0.00 | | 1.05 | 0.91 | 0.00 |
| 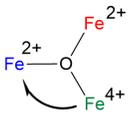 | 0.00 | 1.46 | 1.23 | 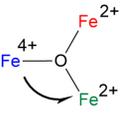 | 0.00 | 1.21 | 1.39 |
| | 0.00 | 1.50 | 1.23 | | 0.00 | 1.19 | 1.39 |
| | 0.00 | 1.50 | 1.23 | | 0.00 | 1.18 | 1.39 |
| | 0.00 | 1.51 | 1.23 | | 0.00 | 1.16 | 1.40 |
| | 0.00 | 1.51 | 1.23 | | 0.00 | 1.19 | 1.39 |
| | 0.00 | 1.46 | 1.23 | | 0.00 | 1.18 | 1.42 |
| | 0.00 | 1.50 | 1.23 | | 0.00 | 1.20 | 1.39 |

15

| 0.00 | 1.50 | 1.23 | | 0.00 | 1.19 | 1.39 |
|---|---|---|---|---|---|---|
| 0.00 | 1.50 | 1.23 | | 0.00 | 1.28 | 1.40 |
| 0.00 | 1.50 | 1.2  | | 0.00 | 1.16 | 1.41 |
| 0.00 | 1.50 | 1.23 | | 0.00 | 1.19 | 1.39 |
| 0.00 | 1.48 | 1.23 | | 0.00 | 1.19 | 1.39 |
| 0.00 | 1.48 | 1.22 | | 0.00 | 1.19 | 1.39 |
| 0.00 | 1.50 | 1.23 | | 0.00 | 1.18 | 1.42 |
| 0.00 | 1.51 | 1.23 | | 0.00 | 1.20 | 1.39 |
| 0.00 | 1.45 | 1.23 | | 0.00 | 1.18 | 1.39 |
| 0.00 | 1.50 | 1.23 | | 0.00 | 1.19 | 1.39 |
| 0.00 | 1.50 | 1.23 | | 0.00 | 1.20 | 1.39 |
| 0.00 | 1.50 | 1.23 | | 0.00 | 1.19 | 1.39 |

# TOC Graphic

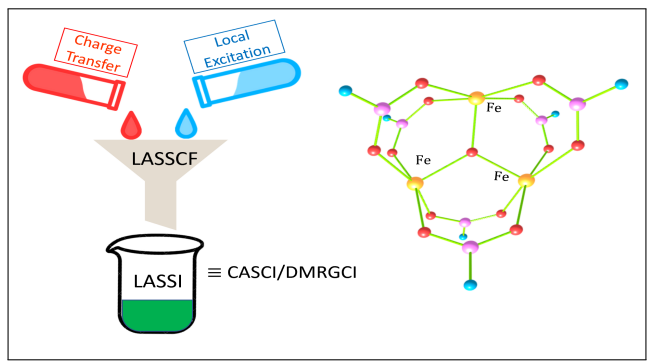


\end{document}